\documentclass[a4paper,fleqn,usenatbib]{mnras}
\usepackage{newtxtext,newtxmath}
\usepackage[T1]{fontenc}
\usepackage{ae,aecompl}

\usepackage{mathtools} 
\usepackage{amsmath} 
\usepackage{amssymb} 
\usepackage{amsfonts} 
\usepackage{mathrsfs} 
\usepackage{graphicx} 
\usepackage{float} 
\usepackage{caption} 
\usepackage{bigstrut} 
\usepackage{tabularx} 
\usepackage{cancel} 
\usepackage{xcolor}
\usepackage{soul}
\usepackage{upgreek}
\usepackage{slashed} 
\usepackage{epstopdf} 
\usepackage{breqn} 
\usepackage[subnum]{cases} 
  {\color{red}}%
  {}

\title[Compressional modes in two-superfluid neutron stars]{Compressional modes in two-superfluid neutron stars with leptonic buoyancy}

\author[P. B. Rau and I. Wasserman]{
Peter B. Rau\thanks{E-mail: pbr44@cornell.edu}
and Ira Wasserman
\\
Cornell Center for Astrophysics and Planetary Science, Cornell University, Ithaca, NY, U.S.A.
}

\date{Accepted 2018 September 5. Received 2018 August 24; in original form 2018 March 1}

\pubyear{2018}

\begin{document}
\label{firstpage}
\pagerange{\pageref{firstpage}--\pageref{lastpage}}
\maketitle


\begin{abstract}
We investigate the compressional modes of cold neutron stars with cores consisting of superfluid neutrons, superconducting protons and normal fluid electrons and muons, and crusts that contain superfluid neutrons plus a normal fluid of (spherical) nuclei and electrons. We develop a two-fluid formalism for the core that accounts for leptonic buoyancy, and an analogous treatment for the crust. We adopt the Cowling approximation, neglecting gravitational perturbations, but include all effects of the background space-time. We introduce a phenomenological, easily-modified nuclear equation of state which contains all of the thermodynamic information required to compute the coupled fluid oscillations, with parameters that are constrained by nuclear physics and the requirement that the maximum mass of a neutron star is $\geq 2M_{\odot}$. Using four parametrizations of this equation of state with nuclear compressibilities $K=230$--$280$ MeV, we calculate the Brunt--V\"{a}is\"{a}l\"{a} frequency due to leptonic buoyancy, and find the corresponding $g$-mode frequencies and eigenfunctions. We find that the WKB approximation reproduces $g$-mode frequencies closely. We examine the dependence of $g$-mode frequencies on stellar mass, nuclear compressibility and strength of neutron-proton entrainment, and compare to previous calculations of $g$-mode frequencies due to leptonic buoyancy. We also compute the $p$-mode spectra, confirming previous findings that the two fluids behave as if uncoupled except the case of large entrainment, and show the existence of nearly resonant mode pairs which could lead to nonlinear $p$-$g$ instabilities even at zero temperature.
\end{abstract}


\section{Introduction}
The first observation of gravitational waves from a binary neutron star merger~\citep{Abbott2017} opens up the possibility of studying neutron stars interiors through tidally-induced phase shifts to gravitational waveforms~\citep{Lackey2015,Agathos2015}. This would allow gravitational wave astronomy to serve as a probe of the equation of state above nuclear density, which is otherwise difficult to study. Low-frequency modes with frequencies swept by the orbital frequency may be resonantly excited by tidal interactions in neutron star-black hole and neutron star-neutron star mergers~\citep{Bildsten1992,Cutler1993,Lai1994,Reisenegger1994,Xu2017,Andersson2018}, causing a phase shift that depends on the exact nature of the excited modes. Low frequency $g$-modes are especially interesting, although the resulting gravitational waveform phase shifts from their resonant excitation will likely be impossible to measure with current-generation detectors unless the merging neutron stars are rapidly rotating or have large radii~\citep{Ho1999,Flanagan2007}. The acoustic $p$-modes are too high in frequency to be resonantly excited themselves, but could participate in nonlinear tidal interactions involving the coupling of the $g$-modes and the $p$-modes which may be observable through a gravitational waveform phase shift~\citep{Weinberg2013,Essick2016}.
 
Potential sources of $g$-modes in neutron stars have been studied for decades. In a normal fluid neutron star, buoyancy arising from temperature gradients~\citep{McDermott1983,Bildsten1995} and the proton fraction gradient~\citep{Reisenegger1992} have been investigated, supporting modes of frequencies $1\sim 100$ Hz. \citet{Lee1995} studied proton fraction gradient $g$-modes in Newtonian stars with superfluid cores, and confirmed a previous calculation by \citet{Lindblom1994} which found two sets of $p$-modes, corresponding to normal fluid and superfluid degree of freedom respectively. Sound speeds for both sets of superfluid neutron star $p$-modes have been calculated by \citet{Epstein1988a}, and this second set of $p$-modes has also been found in a fully relativistic, finite-temperature calculation~\citep{Gualtieri2014}. However, in neutron star cores composed of superfluid neutrons and superfluid-superconducting protons~\citep{Lombardo2001, Page2011}, proton fraction gradients do not lead to $g$-modes unless temperatures are above the neutron critical temperature, estimated to be $\lesssim 10^9$ K ~\citep{Yakovlev1999}, above which electron-neutron coupling~\citep{Bertoni2015} and electron-proton electrostatic coupling will cause both baryon species to move together. $G$-modes due to entropy gradients in superfluid neutron stars were first found by~\citet{Gusakov2013a}, and shortly thereafter they found a new class of $g$-modes resulting from leptonic buoyancy~\citep{Kantor2014} (hereafter KG14). This effect is caused by a gradient in the electron fraction at number densities $\gtrsim 0.13$ fm$^{-3}$, where both electrons and muons coexist in most equations of state. While these leptonic buoyancy $g$-modes were considered in a nonzero temperature star, they were found to exist even in the zero-temperature limit, and their existence was independently confirmed~\citep{Passamonti2016}. A recent paper by~\citet{Yu2017} (hereafter YW17) has computed $g$-mode frequencies and displacement fields arising from leptonic buoyancy in zero temperature neutron star cores using Newtonian gravity, and used their results to study resonant tidal excitation of the modes during neutron star binary inspiral.

We calculate both sets of compressional modes of two-superfluid neutron stars in the zero-temperature approximation, including both the $g$-modes arising due to leptonic buoyancy and the $p$-modes, but with a few crucial differences to previous calculations. Like KG14, we include general relativity and work in the Cowling approximation, neglecting the effects of perturbations to the metric; YW17 used Newtonian gravity but included self gravity perturbations. Secondly, we use a flexible parametrized equation of state (EOS) that allows us to easily adjust the compressibility of the neutron star core, and employ this EOS to calculate the $g$-modes for a range of compressibilities and correspondingly a range of stellar masses and radii. Thirdly, we allow the neutron superfluid to flow into the crust of normal fluid nuclei instead of assuming that the crust is a single normal fluid. We find that this has important implications for the neutron component of the $g$-modes and for both components of the $p$-modes. We compute the displacement fields for the $g$-modes, which KG14 did not report in their initial letter but YW17 did, although the differences in our method mentioned above mean that our modes differ qualitatively and quantitatively. We also use our formalism to compute the $p$-modes in the star like \citet{Lee1995} and \citet{Gualtieri2014}, though only in the zero temperature limit.

In Section~\ref{sec:EoS}, we introduce the parametrized equation of state we use in the core (Section~\ref{subsec:CoreEoS}) and the crust (Section~\ref{subsec:CrustEoS}). In Section~\ref{sec:FluidDynamics} we obtain the equations of motion for the modes and compute the Brunt--V\"{a}is\"{a}l\"{a} frequency due to the muon gradient in the core. The crust-core interface and boundary conditions for the modes, which we find are significant to determining the normal mode displacement fields, are then discussed. Finally, in Section~\ref{sec:NormalModes} we compute the $g$-and $p$-modes, with and without entrainment of the superfluid neutrons and protons in the core, and we make comparisons to previous calculations.

\section{Equation of state}
\label{sec:EoS}

Here we describe the model of the background neutron star that we used for the calculation of the Brunt--V\"{a}is\"{a}l\"{a} frequency in Section~\ref{sec:BVFreq} and the compressional modes in Section~\ref{sec:NormalModes}. Our EOS is based on a relatively simple, parametrized model. We adopt parameters to satisfy constraints near nuclear density $n_{\text{nuc}}=0.16$ fm$^{-3}$ and allow neutron star masses above 2$M_{\odot}$. It is a phenomenological model, but is sufficiently detailed that we can compute all thermodynamic quantities we need to find the normal modes. $\hbar=c=1$ is used throughout.

\subsection{Core equation of state}
\label{subsec:CoreEoS}

We consider an electrically neutral fluid of neutrons, protons, electrons and muons at zero temperature. Its energy density $\rho$ is specified as a function of three variables: baryon number density $n_{\rm b}$, proton fraction $Y$ and electron fraction $f$, where the neutron, proton, electron and muon number densities are respectively $n_{\rm n}=n_{\rm b}(1-Y)$, $n_{\rm p}=n_{\rm b}Y$, $n_{\rm e}=n_{\rm b}Yf$ and $n_{\upmu}=n_{\rm b}Y(1-f)$. We separate the energy density into kinetic and interaction parts $\rho=\rho_{\text{kin}}+\rho_{\text{int}}$; the kinetic part (including the rest mass) is given by
\begin{equation}
\rho_{\text{kin}}(n_{\rm b},Y,f)=\frac{p_{\rm Fe}^4}{4\pi^2}+\sum_{j={\rm n,p,}\upmu}\frac{m_j^4}{\pi^2}\phi\left(\frac{p_{{\rm F}j}}{m_j}\right)
\label{eq:KineticEnergyDensity}
\end{equation}
for Fermi momenta $p_{{\rm F}j}=(3\pi^2n_j)^{1/3}$ and bare mass $m_j$ of particle species $j$, and where
\begin{equation}
\phi(x)=\frac{x^3}{4}\sqrt{x^2+1}+\frac{x}{8}\sqrt{x^2+1}-\frac{1}{8}\text{arsinh}(x).
\end{equation}
We have assumed the electrons are ultrarelativistic and ignore the difference between the proton and neutron mass, assuming $m_{\rm n}=m_p=m_{\rm N}$. Although we use the bare nucleon mass, we assume that $\rho_{\text{int}}$ includes effective mass corrections adequately. The interaction energy density $\rho_{\text{int}}(n_{\rm b},Y)$ employed is based on that of~\citet{Hebeler2013}, but with a different form for the symmetry penalty term:
\begin{align}
\rho_{\text{int}}(n_{\rm b},Y)={}&n_{\text{nuc}}E_{\rm S}\frac{\overline{n}^2+f_{\rm S}\overline{n}^{\gamma_{\rm S}+1}}{1+f_{\rm S}} \nonumber
\\ & +n_{\text{nuc}}E_{\rm A}\overline{n}^2\left(\frac{\overline{n}+\overline{n}_0}{1+\overline{n}_0}\right)^{\gamma_{\rm A}-1}(1-2Y)^2,
\label{eq:NucleonInteractionEnergyDensity}
\end{align}
where $\overline{n}=n_{\rm b}/n_{\text{nuc}}$ and $\overline{n}_0$ is a characteristic number density. For $\overline{n}\ll\overline{n}_0$, the symmetry penalty term is quadratic in $\overline{n}$, as the energy per baryon should be linear in the density at low densities.
    
The requirements of $-16$ MeV per baryon binding energy and zero pressure for symmetric nuclear matter at nuclear density constraint the parameters $E_{\rm S}$, $\gamma_{\rm S}$ and $f_{\rm S}$, while experimental measurements like those used in constructing Figure~6 of~\citet{Lattimer2016} constrain $E_{\rm A}$, $\gamma_{\rm A}$ and $\overline{n}_0$. Another constraint is that the EOS must allow a maximum mass $\geq 2M_{\odot}$~\citep{Antoniadis2013}, though this can be adjusted upward to allow for higher masses if required by future observations. We consider four possible parameter choices PC1--PC4, differing in the values of $\gamma_{\rm S}$ and $f_{\rm S}$, with each choice corresponding to a value of the nuclear compressibility parameter $K=9(\partial^2(\rho/n_{\rm b})/\partial\overline{n}^2)|_{\overline{n}=1,Y=1/2}$. These are listed in Table~\ref{tab:EOSParameters}. For these choices of $E_{\rm A}$, $\gamma_{\rm A}$ and $\overline{n}_0$, the symmetry energy $S_{\rm v}=31.73$ MeV and density derivative $L=60.32$ MeV, within the $1\sigma$ confidence region of Figure~6 of~\citet{Lattimer2016}. Three of the chosen values of $K$ are within the $240\pm20$ MeV confidence range cited in~\citet{Lattimer2016}, with $K=220$ MeV not being used due to not allowing a $2M_{\odot}$ star to exist in our EOS. The $K=280$ MeV parametrization represents a causal limit i.e. the sound speed equals the speed of light for central densities just beyond that which has the maximum mass for this parametrization. While the EOS is flexible, we found that it was difficult to obtain a maximum mass greater than $2.2M_{\odot}$, and also found that adjusting the parameters $E_{\rm A}$, $\gamma_{\rm A}$ and $\overline{n}_0$ had only a small effect on the nuclear compressibility, so these parameters were fixed for all four parameter sets.

\begin{table}
	\caption{Different parametrizations of core equation of state and corresponding nuclear compressibility $K$, maximum mass $M_{\text{max}}$, radius at maximum mass $R_{\text{max}}$, central baryon number density for the maximum mass star $n_{b,\text{cntr},\text{max}}$ and the baryon number density at which the sound speed equals the speed of light $n_{b,\text{cl}}$.}
	  \centering
    \begin{tabular}{|c|c|c|c|c|}
    \hline
	\multicolumn{1}{|c|}{} & \textbf{PC1} & \textbf{PC2} & \textbf{PC3} & \textbf{PC4} \\ 
    \hline
    $E_{\rm S}$ (MeV) & -37.8 & -37.8 & -37.8 & -37.8 \\
	\hline
	$\gamma_{\rm S}$ & 1.31 & 1.356 & 1.452 & 1.547 \\
	\hline  
	$f_{\rm S}$ & -0.667 & -0.634 & -0.577 & -0.530 \\
	\hline  
	$E_{\rm A}$ (MeV) & 19.9 & 19.9 & 19.9 & 19.9 \\
	\hline  
	$\gamma_{\rm A}$ & 0.61 & 0.61 & 0.61 & 0.61 \\
	\hline  
	$\overline{n}_0$ (MeV) & 0.05 & 0.05 & 0.05 & 0.05 \\
	\hline  
	$K$ (MeV) & 230 & 240 & 260 & 280 \\
	\hline
	$M_{\text{max}}/M_{\odot}$ & 2.01 & 2.05 & 2.15 & 2.24 \\
	\hline
	$R_{\text{max}}$ (km) & 10.23 & 10.34 & 10.62 & 10.88 \\
	\hline
	$n_{{\rm b},\text{cntr},\text{max}}/n_{\text{nuc}}$ & 7.43 & 7.22 & 6.73 & 6.32 \\
	\hline
	$n_{{\rm b},\text{cl}}/n_{\text{nuc}}$ & 9.8 & 8.9 & 7.5 & 6.4 \\
	\hline        
    \end{tabular}
    \label{tab:EOSParameters}
\end{table}
    
\begin{figure}
\centering
\includegraphics[width=1.0\columnwidth]{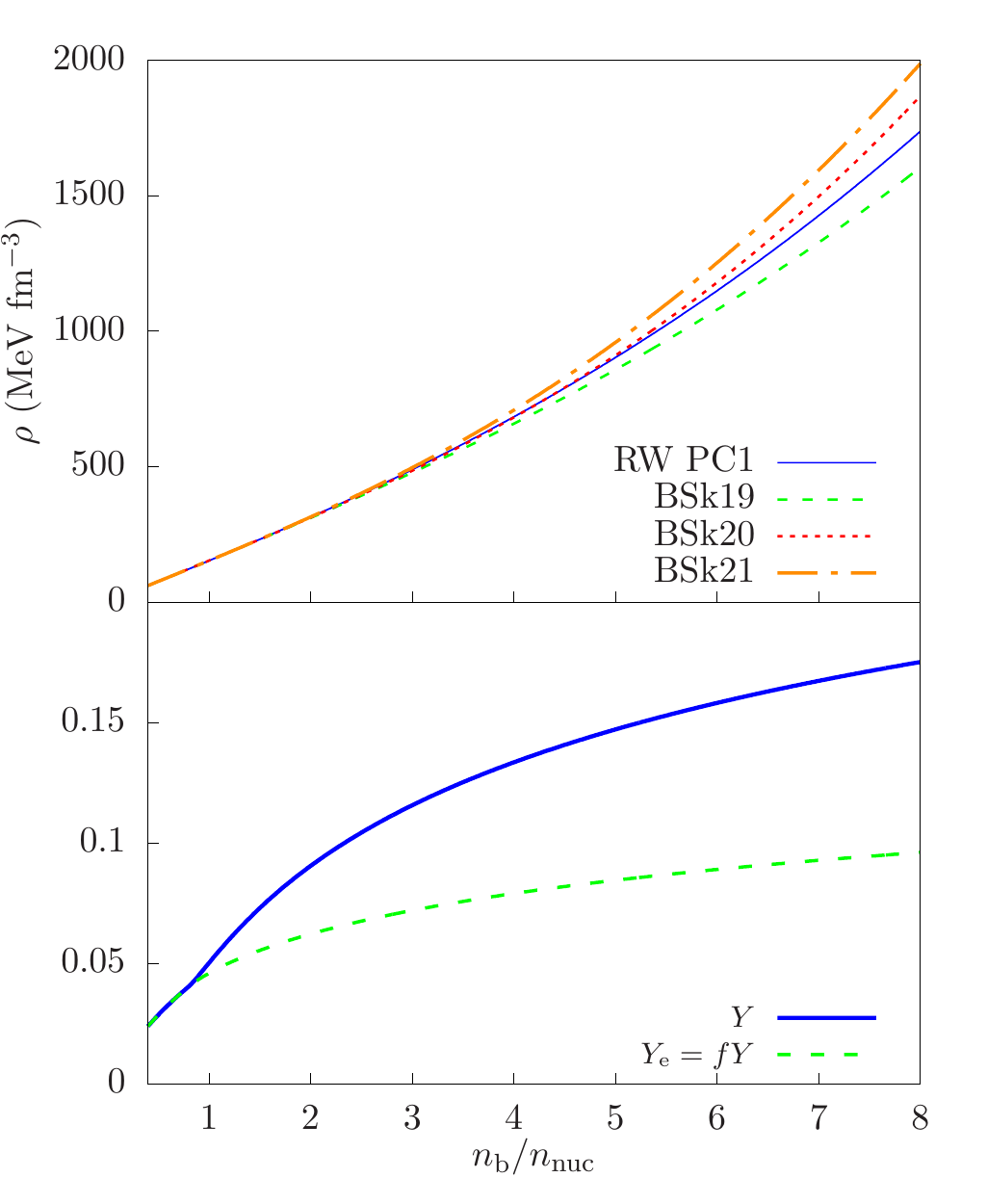}
\caption{Top: Energy density $\rho$ as a function of $n_{\rm b}/n_{\text{nuc}}$ in the core from this paper (RW with PC1 parameters) and the BSk19--BSk21 EOS from~\citet{Potekhin2013}. Bottom: Proton fraction $Y=n_{\rm q}/n_{\rm b}$ and electron fraction $Y_{\rm e}=fY=n_{\rm e}/n_{\rm b}$ in the core for the RW EOS with PC1 parameters. $Y=Y_{\rm e}$ below the muon threshold density $n_{\rm b}/n_{\text{nuc}}=0.8$. }
\label{fig:CoreEoS}
\end{figure}

The pressure is specified by
\begin{equation}
P=n_{\rm b}\frac{\partial \rho}{\partial n_{\rm b}}-\rho
\end{equation}
and the chemical potential by
\begin{equation}
\mu=\frac{\partial \rho}{\partial n_{\rm b}}.
\end{equation}
The individual chemical potentials are calculated using
\begin{equation}
\mu_x=\frac{\partial \rho}{\partial n_x},\quad x={\rm n,p,e},\upmu.
\end{equation}
The background star is assumed to be in beta equilibrium, implying
\begin{equation}
\mu_{\rm n}=\mu_{\rm p}+\mu_{\rm e}=\mu_{\rm p}+\mu_{\upmu} \Rightarrow \mu_{\rm e}=\mu_{\upmu}.
\end{equation}
We find that muons first appear at $n_{\rm b}=0.8n_{\text{nuc}}$ for all four EOS parametrizations that we considered. In Figure~\ref{fig:CoreEoS} (top), we compare our $\rho(n_{\rm b})$ in the core to the BSk19--BSk21 EOSs from~\citet{Potekhin2013}, finding that ours is in good agreement with all three of their EOSs in the lower half of the density range and with the BSk19 and BSk20 EOS in the higher density region. We also plot the proton fraction $Y$ and $Y_{\rm e}=fY$ as functions of $n_{\rm b}$ in the core (bottom).

\subsection{Crust equation of state}
\label{subsec:CrustEoS}

In the inner crust between neutron drip at $n_{\rm b}\sim 10^{11}$ g/cm$^3$ and the transition to the core at $\sim 10^{14}$ g/cm$^3$ ($n_{\rm b}\sim 0.5n_{\text{nuc}}$), neutron stars are expected to consist of neutron-rich nuclei surrounded by a dripped neutron gas and a pervasive ultrarelativistic electron gas. Below neutron drip, the outer crust, consisting only of nuclei and an electron gas, is included using the BPS EOS~\citep{Baym1971a}. It is, however, neglected when computing the oscillation modes as it constitutes less than a hundredth of a percent of the star by mass, thus having a negligible effect on the bulk oscillation modes. The effects of this omission are briefly discussed at the conclusion of Section~\ref{sec:BCs}.

Following~\citet{Baym1971} and~\citet{Haensel2001b}, we consider a liquid drop-type model with spherical nuclei of radius $r_{\rm n}$ inside spherical unit cells of radius $r_{\rm c}$. We do not model exotic shapes or nuclear pasta~\citep{Ravenhall1983,Hashimoto1984,Watanabe2003} at this stage, and we allow the proton and nucleon numbers $Z$ and $A$ to vary continuously. The density of neutrons outside the nuclei is $n_{\rm n,o}$, while the nuclei themselves have baryon density $n_{\rm i}$. The neutron and proton densities inside the nuclei are $n_{\rm n,i}=(1-Y)n_{\rm i}$ and $n_{\rm p,i}=Yn_{\rm i}$ respectively, where in the crust $Y=Z/A$ is the proton fraction of the nuclei and $Z$ and $A$ are defined to include only baryons inside the nuclei. The electron number density is fixed by electric charge neutrality to be equal to the average proton number density over the cell, so
\begin{equation}
n_{\rm e} = wn_{\rm p,i}=wn_{\rm i}Y,
\end{equation}
where $w=(r_{\rm n}/r_{\rm c})^3$ is the fraction of the volume of each cell occupied by the nucleus and $r_{\rm n}=(3A/4\pi n_{\rm i})^{1/3}$. We also define the number density of nuclei $n_{\rm n}$, which is given in terms of the cell radius by
\begin{equation}
n_{\rm n} = \frac{3}{4\pi r_{\rm c}^3},
\end{equation}
so the total baryon number density $n_{\rm b}$ is given by
\begin{equation}
n_{\rm b} = An_{\rm n}+(1-w)n_{\rm n,o}.
\end{equation}

Later, we discuss the fluid oscillations in the crust in terms of the macroscopic motion of two fluids: a free neutron superfluid and a normal fluid of nuclei. In  the fluid equations, we use the mean density of free neutrons outside the nuclei $n_{\rm f}\equiv(1-w)n_{\rm n,o}$ and the mean density of nuclear baryons $n_{\rm c}\equiv An_{\rm n}$. In terms of these densities, the total baryon number density is
\begin{equation}
n_{\rm b}=n_{\rm f}+n_{\rm c};
\end{equation}
note too that $w=n_{\rm c}/n_{\rm i}$.

We write the energy density for the inner crust in terms of the five variables $n_{\rm f}$, $n_{\rm c}$, $n_{\rm i}$, $A$ and $Y$. The energy density for the inner crust has five components: bulk energy densities for the nuclei $\rho_{i,\text{bulk}}$, surrounding neutron gas $\rho_{{\rm o},\text{bulk}}$, and electron gas $\rho_{\rm e}$, Coulomb energy density $\rho_{\text{Coul}}$ including the self-energy of the nuclei and the lattice energy, and surface energy density $\rho_{\text{surf}}$. The bulk energy density for nuclear matter, the neutron gas and the electron gas have the same form as in the core, discussed in the previous section. We then have
\begin{align}
\rho(n_{\rm f},n_{\rm c},A,Y,n_{\rm i}) ={}& w\rho_{{\rm i},\text{bulk}}(n_{\rm i},Y)+(1-w)\rho_{{\rm o},\text{bulk}}(n_{\rm f}/(1-w))\nonumber \\
{}&+\rho_{\rm e}(Yn_{\rm c})+n_{\rm n}E_{\text{Coul}}(n_{\rm c},n_{\rm i},A,Y)\nonumber \\
{}&+n_{\rm n}E_{\text{surf}}(n_{\rm i},A,Y), \label{eq:CrustEnergyDensity}
\end{align}
where
\begin{align}
{}&\rho_{{\rm i},\text{bulk}}(n_{\rm i},Y)=\frac{m_{\rm N}^4}{\pi^2}\left[\phi\left(\frac{p_{\rm Fi}}{m_{\rm N}}Y^{1/3}\right)+\phi\left(\frac{p_{\rm Fi}}{m_{\rm N}}(1-Y)^{1/3}\right)\right] \nonumber \\
&\quad +n_{\text{nuc}}E_{\rm S}\frac{\overline{n}_{\rm i}^2+f_{\rm S}\overline{n}_{\rm i}^{\gamma_{\rm S}+1}}{1+f_{\rm S}}+n_{\text{nuc}}E_{\rm A}\overline{n}_{\rm i}^2\left(\frac{\overline{n}_{\rm i}+\overline{n}_{\rm o}}{1+\overline{n}_{\rm o}}\right)^{\gamma_{\rm A}-1}(1-2Y)^2, \\
{}&\rho_{{\rm o},\text{bulk}}(n_{\rm n,o})=\frac{m_{\rm N}^4}{\pi^2}\phi\left(\frac{p_{\rm Fo}}{m_{\rm N}}\right)+n_{\text{nuc}}E_{\rm S}\frac{\overline{n}_{\rm n,o}^2+f_{\rm S}\overline{n}_{\rm n,o}^{\gamma_{\rm S}+1}}{1+f_{\rm S}} \nonumber \\
&\qquad +n_{\text{nuc}}E_{\rm A}\overline{n}^2_{\rm n,o}\left(\frac{\overline{n}_{\rm n,o}+\overline{n}_{\rm o}}{1+\overline{n}_{\rm o}}\right)^{\gamma_{\rm A}-1}, \\
{}&\rho_{\rm e}(Yn_{\rm c}=n_{\rm e})=\frac{(3\pi^2n_{\rm e})^{4/3}}{4\pi^2}, \\
{}&E_{\text{Coul}}(n_{\rm c},n_{\rm i},A,Y)=\frac{16}{15}(\pi Yn_{\rm i}e)^2r_{\rm n}^5\left[1-\frac{3}{2}w^{1/3}+\frac{1}{2}w\right], \\
{}&E_{\text{surf}}(n_{\rm i},A,Y)=4\pi r_{\rm n}^2\sigma_{\rm s}(Y),
\end{align}
where $p_{\rm Fi}=(3\pi^2n_{\rm i})^{1/3}$, $n_{\rm n,o}=n_{\rm f}/(1-w)$, $p_{\rm Fo}=(3\pi^2n_{\rm n,o})^{1/3}$, $\overline{n}_{\rm i}=n_{\rm i}/n_{\text{nuc}}$, $\overline{n}_{\rm n,o}=n_{\rm n,o}/n_{\text{nuc}}$ and $\sigma_{\rm s}$ denotes the nuclear surface energy. We assume that the electrons are relativistic down to neutron drip and ignore the neutron-proton mass difference here.

Following~\citet{Ravenhall1983a} and~\citet{Lattimer1985}, we take the surface energy $\sigma_s$ to be a function only of the proton fraction $Y$ at zero temperature, and use the parametrization
\begin{equation}
\sigma_{\rm s}(Y)=\frac{\sigma_0(2^{\alpha+1}+\beta)}{Y^{-\alpha}+(1-Y)^{-\alpha}} .
\label{eq:SurfaceEnergy}
\end{equation}
We selected parameters $\sigma_0$, $\alpha$, $\beta$ which give a approximately constant proton number $Z\approx 40$ throughout the density range of the inner crust, as is found in more detailed calculations of the inner crust equation of state \citep{Douchin2000,Onsi2008,Pearson2012,Potekhin2013}:
\begin{align*}
&\sigma_0 = 1.4\text{ MeV/fm}^2, \\
&\alpha = 3, \\
&\beta = 24.
\end{align*}
$\alpha$ and $\beta$ are close to the corresponding parameters in~\citet{Ravenhall1983a}, but $\sigma_0$ is $\approx 50$\% larger than its corresponding parameter.

For a general change of state, the change in the energy density is
\begin{align}
d\rho={}&\left.\frac{\partial\rho}{\partial A}\right|_{n_{\rm f},n_{\rm c},n_{\rm i},Y}{\rm d}A+\left.\frac{\partial\rho}{\partial Y}\right|_{n_{\rm f},n_{\rm c},n_{\rm i},A}{\rm d}Y \nonumber \\
{}&+\left.\frac{\partial\rho}{\partial n_{\rm i}}\right|_{n_{\rm f},n_{\rm c},A,Y}{\rm d}n_{\rm i}+\left.\frac{\partial\rho}{\partial n_{\rm c}}\right|_{n_{\rm f},n_{\rm i},A,Y}{\rm d}n_{\rm c} \nonumber \\
{}&+\left.\frac{\partial\rho}{\partial n_{\rm f}}\right|_{n_{\rm c},n_{\rm i},A,Y}{\rm d}n_{\rm f}.
\end{align}
At fixed $n_{\rm b}$, ${\rm d}n_{\rm c}+{\rm d}n_{\rm f}=0$, so this becomes
\begin{align}
&d\rho=\left.\frac{\partial\rho}{\partial A}\right|_{n_{\rm f},n_{\rm c},n_{\rm i},Y}{\rm d}A+\left.\frac{\partial\rho}{\partial Y}\right|_{n_{\rm f},n_{\rm c},n_{\rm i},A}{\rm d}Y \nonumber \\
&+\left.\frac{\partial\rho}{\partial n_{\rm i}}\right|_{n_{\rm f},n_{\rm c},A,Y}{\rm d}n_{\rm i}+\left(\left.\frac{\partial\rho}{\partial n_{\rm c}}\right|_{n_{\rm f},n_{\rm i},A,Y}-\left.\frac{\partial\rho}{\partial n_{\rm f}}\right|_{n_{\rm c},n_{\rm i},A,Y}\right){\rm d}n_{\rm c}.
\end{align}
The ``nuclear virial theorem'' \citep{Haensel2001b} and pressure balance correspond to the conditions
\begin{equation}
\left.\frac{\partial\rho}{\partial Y}\right|_{n_{\rm f},n_{\rm c},n_{\rm i},A}=
\left.\frac{\partial\rho}{\partial n_{\rm i}}\right|_{n_{\rm f},n_{\rm c},A,Y} = 0, \label{eq:NVTandPB}
\end{equation}
respectively, while the condition that there is no energy associated with exchanging neutrons between the nuclei and the external free neutron gas (henceforth the ``exchange condition'') is
\begin{equation}
\left.\frac{\partial\rho}{\partial n_{\rm c}}\right|_{n_{\rm c},n_{\rm i},A,Y}-\left.\frac{\partial\rho}{\partial n_{\rm f}}\right|_{n_{\rm c},n_{\rm i},A,Y}-\frac{Y}{n_{\rm c}}\left.\frac{\partial\rho}{\partial Y}\right|_{n_{\rm f},n_{\rm c},n_{\rm i},A}=0 \label{eq:Exchange1}
\end{equation}
since proton density $Yn_{\rm c}$ is unchanged by exchange of neutrons. Beta equilibrium is simply
\begin{equation}
\left.\frac{\partial\rho}{\partial Y}\right|_{n_{\rm f},n_{\rm c},n_{\rm i},A}=0, \label{eq:BetaEq}
\end{equation}
so Eq.~(\ref{eq:Exchange1}) becomes
\begin{equation}
\left.\frac{\partial\rho}{\partial n_{\rm c}}\right|_{n_{\rm c},n_{\rm i},A,Y}-\left.\frac{\partial\rho}{\partial n_{\rm f}}\right|_{n_{\rm c},n_{\rm i},A,Y}=0. \label{eq:Exchange2}
\end{equation}
Imposing the four conditions Eqs.~(\ref{eq:NVTandPB},\ref{eq:BetaEq},\ref{eq:Exchange2}), we can determine the values of $n_{\rm f}$, $n_{\rm c}$, $n_{\rm i}$, $A$ and $Y$ at each $n_{\rm b}$ and thus compute at each $n_{\rm b}$ the energy density $\rho$ and pressure $P$, which is given by
\begin{equation}
P = P_{\rm e}+P_{\text{Coul}}+P_{{\rm o},\text{bulk}}=\frac{1}{3}\rho_{\rm e}+\frac{n_{\rm c}^2}{A}\frac{\partial E_{\text{Coul}}}{\partial n_{\rm c}}+P_{{\rm o},\text{bulk}}.
\label{eq:CrustPressure}
\end{equation}
We will also require the chemical potentials for each fluid $\mu_{\rm f}$ and $\mu_{\rm c}$, given by
\begin{align}
\mu_{\rm f}=\left.\frac{\partial\rho}{\partial n_{\rm f}}\right|_{n_{\rm c},n_{\rm i},A,Y}={}&(1-w)\frac{\partial\rho_{\text{bulk},{\rm o}}}{\partial n_{\rm n,o}}\frac{\partial n_{\rm n,o}}{\partial n_{\rm f}}=\mu_{\rm n,o}, \label{eq:MuFUnsimplified} \\
\mu_{\rm c}=\left.\frac{\partial\rho}{\partial n_{\rm c}}\right|_{n_{\rm f},n_{\rm i},A,Y}={}&Y\mu_{\rm e}+\frac{P_{{\rm o},\text{bulk}}+\rho_{\text{bulk},i}}{n_{\rm i}} \nonumber \\ 
{}&+\frac{16}{15}(\pi Yn_{\rm i}e)^2\frac{Y}{A}r_{\rm n}^5\left[3-5w^{1/3}+2w\right]. \label{eq:MuCUnsimplified}
\end{align}
Using the four equilibrium conditions, Eqs.~(\ref{eq:MuFUnsimplified}--\ref{eq:MuCUnsimplified}) can be used to show that $\mu_{\rm c}=\mu_{\rm f}$ in equilibrium.

For this inner crust equation of state, neutron drip occurs between $2.7-2.8\times 10^{11}$ g/cm$^3$ or $n_{\rm b}=0.00103-0.00106n_{\text{nuc}}$, depending on the parametrization of the nuclear physics. The core and crust equations of state must be joined when the pressure and chemical potentials of each are equal. This occurs at different densities for each EOS parametrization described in Table~\ref{tab:EOSParameters}, with the pressure, chemical potential and densities in the core and crust at the transition for each parametrization listed in Table~\ref{tab:CCTransition}. 

\begin{table}
	\caption{Pressure $P$, chemical potential $\mu$ and baryon number density $n_{\rm b}$ at the core ($+$) and crust ($-$) sides of the crust-core transition for each EOS parametrization employed in this paper. The size of the density jump as a percentage of the baryon number density at the transition is also listed.}
	  \centering
    \begin{tabular}{|c|c|c|c|c|}
    \hline
	\multicolumn{1}{|c|}{} & \textbf{PC1} & \textbf{PC2} & \textbf{PC3} & \textbf{PC4} \\ 
    \hline
    $P$ (MeV/fm$^3$) & 0.215 & 0.225 & 0.239 & 0.249 \\
	\hline
	$\mu/m_{\rm N}$ & 1.0108 & 1.0112 & 1.0117 & 1.0121 \\
	\hline  
	$n_{\rm b}^+/n_{\text{nuc}}$ & 0.399 & 0.413 & 0.437 & 0.458 \\
	\hline  
	$n_{\rm b}^-/n_{\text{nuc}}$ & 0.394 & 0.407 & 0.431 & 0.451 \\
	\hline   
	$\Delta n_{\rm b}/n_{\rm b}$ (\%) & 1.2 & 1.4 & 1.3 & 1.5 \\
	\hline      
    \end{tabular}
    \label{tab:CCTransition}
\end{table}

\begin{figure}
\centering
\includegraphics[width=1.0\columnwidth]{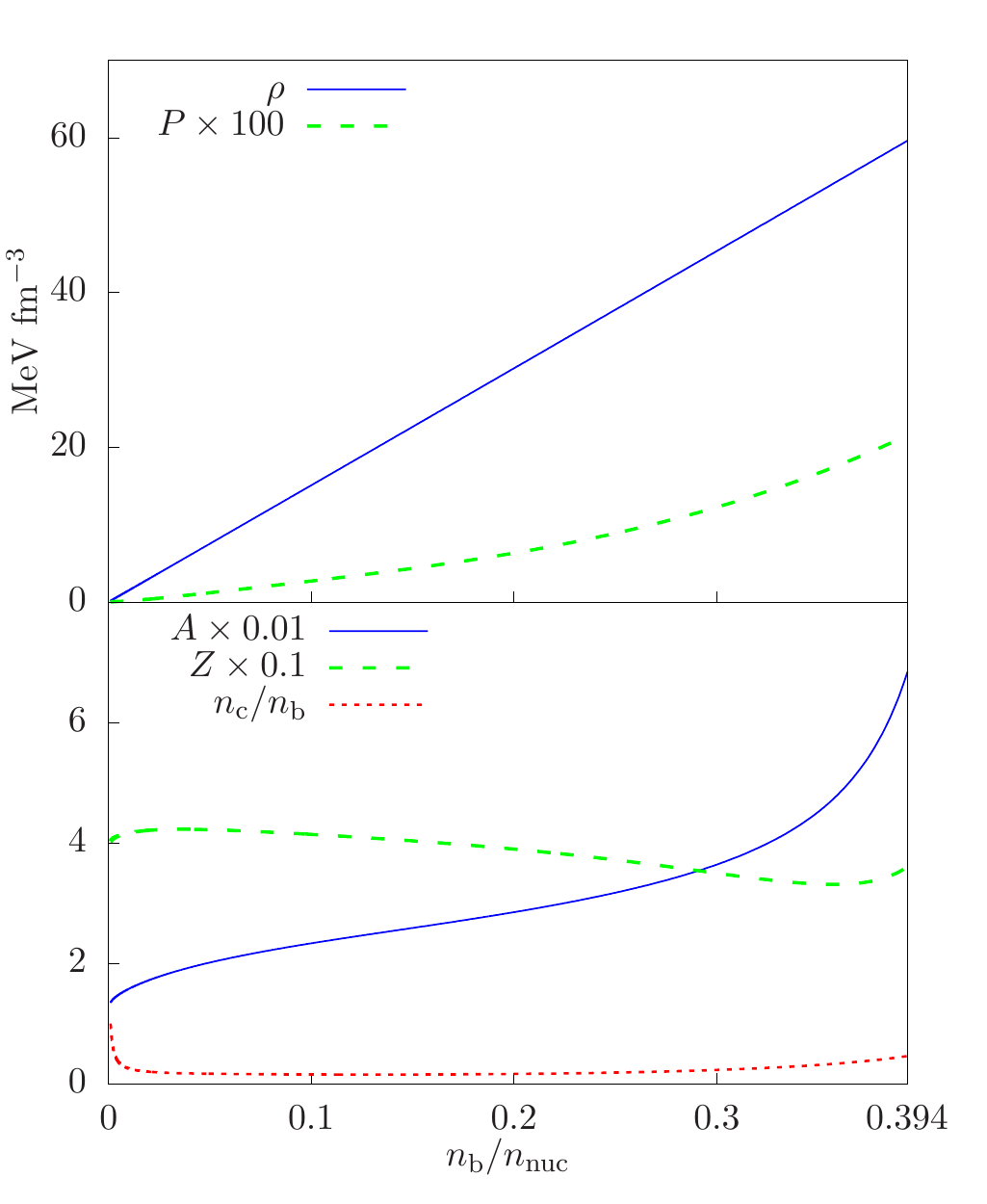}
\caption{Top: Mass-energy density $\rho$ and pressure $P$ as functions of $n_{\rm b}/n_{\text{nuc}}$ across the density range of the inner crust for the RW EOS with PC1 parametrization. Bottom: Nucleon number $A$, proton number $Z$ and the ratio $0<n_{\rm c}/n_{\rm b}<1$ as functions of $n_{\rm b}/n_{\text{nuc}}$ across the density range of the inner crust for the same EOS. The maximum density in the crust before the transition to the core is $n_{\rm b}/n_{\text{nuc}}=0.394$ for the PC1 parametrization.}
\label{fig:CrustEOS}
\end{figure}

The energy density and pressure for our EOS in the crust (Figure~(\ref{fig:CrustEOS}), top) are within 10--20\% of those found in more detailed calculations such as~\citet{Pearson2012} and~\citet{Potekhin2013}. The bottom panel shows $A$, $Z$ and $n_{\rm c}/n_{\rm b}$ as functions of $n_{\rm b}/n_{\text{nuc}}$ in the crust. The values of $Z$ closely match those of~\citet{Ravenhall1972}, including the dip downward just before the transition to the core. Our values of $n_{\rm c}/n_{\rm b}$ are in agreement with~\citet{Kobyakov2016}; our values of $A$ are typically greater than theirs by a factor of $1.5$, though the difference increases as the crust-core transition is approached, while our values are typically a factor of $4$ lower than those of Pearson et al. 

\begin{figure}
\centering
\includegraphics[width=1.0\columnwidth]{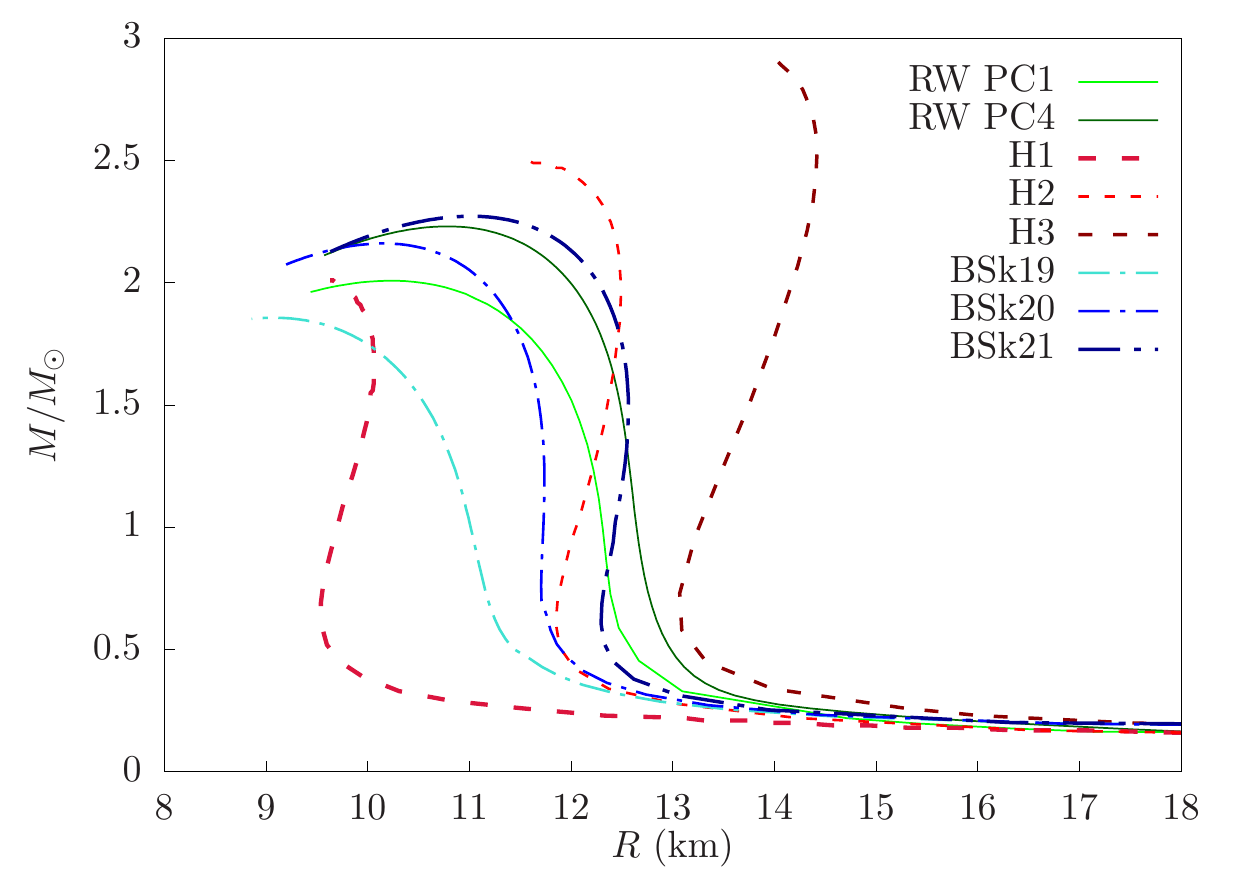}
\caption{Neutron star mass-radius plot for the EOS described in this paper with two parametrizations as given by Table~\ref{tab:EOSParameters} (RW PC1 and RW PC4) and three equations of state of increasing stiffness from both~\citet{Hebeler2013} (denoted H1, H2, H3) and~\citet{Potekhin2013} (denoted BSk19, BSk20, BSk21).}
\label{fig:MvsR}
\end{figure}

Figure~\ref{fig:MvsR} compares the mass-radius relation for two different parametrizations of the two-part EOS described here (RW) to a few other representative neutron star equations of state. The RW EOS uses the BPS EOS~\citep{Baym1971a} for densities below neutron drip. The radius, radius at neutron drip and central density for each EOS parametrization and stellar mass used in the rest of this paper are described in Table~\ref{tab:StellarModelProperties}.

\begin{table}
	\caption{List of radius $R$, radius at neutron drip $R_{\text{ND}}$, and central density $n_{b,\text{cntr}}$ for each stellar mass and parametrization choice employed in the calculation of compressional mode frequencies in this paper.}
	  \centering
    \begin{tabular}{|c|c|c|c|c|}
    \hline
	 \textbf{PC1}  & $\mathbf{1.2M_{\odot}}$ & $\mathbf{1.4M_{\odot}}$ & $\mathbf{1.7M_{\odot}}$ & $\mathbf{2M_{\odot}}$ \\ 
    \hline
   	$R$ (km) & 12.24 & 12.11 & 11.75 & 10.54 \\
	\hline
	$R_{\text{ND}}$ (km) & 11.72 & 11.70 & 11.47 & 10.39 \\
	\hline  
	$n_{\rm b,\text{cntr}}/n_{\text{nuc}}$ & 2.73 & 3.17 & 4.07 & 6.67 \\
	\hline  
	\textbf{PC2} & $\mathbf{1.2M_{\odot}}$ & $\mathbf{1.4M_{\odot}}$ & $\mathbf{1.7M_{\odot}}$ & $\mathbf{2M_{\odot}}$ \\
	\hline	
	$R$ (km) & 12.31 & 12.21 & 11.90 & 11.06 \\
	\hline
	$R_{\text{ND}}$ (km) & 11.78 & 11.79 & 11.62 & 10.89 \\
	\hline  
	$n_{\rm b,\text{cntr}}/n_{\text{nuc}}$ & 2.66 & 3.05 & 3.86 & 5.61 \\
	\hline	
	\textbf{PC3} & $\mathbf{1.2M_{\odot}}$ & $\mathbf{1.4M_{\odot}}$ & $\mathbf{1.7M_{\odot}}$ & $\mathbf{2M_{\odot}}$ \\ 
	\hline	
	$R$ (km) & 12.48 & 12.43 & 12.23 & 11.71 \\
	\hline
	$R_{\text{ND}}$ (km) & 11.94 & 11.99 & 11.92 & 11.50  \\
	\hline  
	$n_{\rm b,\text{cntr}}/n_{\text{nuc}}$ & 2.49 & 2.82 & 3.46 & 4.55 \\
	\hline 	
	\textbf{PC4} & $\mathbf{1.2M_{\odot}}$ & $\mathbf{1.4M_{\odot}}$ & $\mathbf{1.7M_{\odot}}$ & $\mathbf{2M_{\odot}}$ \\ 
	\hline	
	$R$ (km) & 12.63 & 12.62 & 12.50 & 12.14 \\
	\hline
	$R_{\text{ND}}$ (km) & 12.07 & 12.17 & 12.17 & 11.92 \\
	\hline  
	$n_{\rm b,\text{cntr}}/n_{\text{nuc}}$ & 2.35 & 2.64 & 3.16 & 3.96 \\
	\hline       
    \end{tabular}
    \label{tab:StellarModelProperties}
\end{table}

\section{Fluid dynamics}
\label{sec:FluidDynamics}

\subsection{Two-fluid formalism in the core}

We now derive the equations of motion for the perturbations of a two-superfluid neutron star that we use to compute its normal modes. We first consider the core, and the same model is generalized to the crust in Sections~(\ref{subsec:CrustEoM}) and~(\ref{subsec:SFCrustEoM}). We work at zero temperature, so there are no normal fluid neutron or proton components, and include general relativity, but work in the Cowling approximation and so ignore perturbations of the metric.

In a core composed of superfluid neutrons and protons and normal fluid electrons and muons, the leptons will move along with the protons since the plasma frequency $\sim10^{22}$ s$^{-1}$ is much greater than the frequencies of the compressional modes. We thus have two independently-moving fluids-- a neutron superfluid and charged fluid. This differs from the commonly-chosen separation of the fluid into normal and superfluid components, though the formulation used here has a number of advantages which illuminate the underlying physics. First, it makes clear the role of the leptonic buoyancy, which exists only in the charged fluid. It also reveals the significance of thermodynamic coupling and entrainment between the two fluids, with the equations describing the motion of the fluids becoming completely uncoupled if these quantities are zero as is demonstrated at the end of this section. At densities above $2$--$3n_{\text{nuc}}$, the $s$-wave pairing energy gap for protons may go to zero~\citep{Zhou2004,Baldo2007}, meaning that the proton fluid can be a normal fluid in the inner core, but the equations of motion for the charged fluid will remain unchanged in this case, and the charged fluid and neutrons would still behave as two separately-moving fluids as long as the neutrons remain superfluid. Note that both methods should be equivalent, and the normal fluid and superfluid displacement modes can be reconstructed from taking the appropriate superposition of the neutron and charged fluid modes.

We assume that neutrons remain superfluid throughout the core; calculations summarized in Fig. 2 of~\citet{Gezerlis2014} support the idea that the neutron gap does not vanish at any density below at least $4.2n_{\text{nuc}}$,  which includes all neutron stars less massive than about $1.7M_{\odot}$ for our adopted equation of state. Moreover, these calculations suggest that neutron superfluidity would be maintained out to the crust-core boundary for core temperatures below $\simeq 3\times 10^8$K; the model used in KG14 is similar. If neutrons become normal somewhere inside the core, their coupling to electrons~\citep{Bertoni2015} suffices to merge the two fluids into a single fluid, irrespective of whether the protons are superfluid. Thus, if neutrons were entirely normal throughout the core, then $g$-modes would arise from a combination of the leptonic buoyant force associated with the gradient of $f$ and the buoyant force associated with the gradient of $Y$~\citep{Reisenegger1992}, but overall  their frequencies would be lower than when neutrons are superfluid; see Section~\ref{sec:BVFreq}, especially Eqs.~(\ref{eq:NFBruntVaisalaFrequency}--\ref{eq:YGradientBruntVaisalaFrequency}). However, if the neutrons are only normal deep inside the core of the neutron star, then the $g$-modes arising from leptonic buoyancy, which is only substantial near the outer boundary of the core, would be largely unaffected. Thus, the $g$-mode frequency spectrum is, in principle, a probe of neutron superfluidity in the cores of neutron stars.

We specify the neutron superfluid four-velocity $u^{\mu}_{\rm n}$ and number density $n_{\rm n}$, and charged fluid four-velocity $u^{\mu}_{\rm q}$ and number density $n_{\rm q}=n_{\rm p}=n_{\rm e}+n_{\upmu}$. We can rewrite the energy density $\rho$ as a function of $n_{\rm n}$, $n_{\rm q}$ and the electron fraction $f=n_{\rm e}/n_{\rm q}$;
\begin{equation}
\rho(n_{\rm n},n_{\rm q},f)=\rho_{\text{nuc}}(n_{\rm n},n_{\rm q})+\rho_{\rm e}(n_{\rm q}f)+\rho_{\upmu}(n_{\rm q}(1-f)),
\end{equation}
where $\rho_{\text{nuc}}$ includes both the kinetic and interaction contributions relating to the nucleons. This gives two chemical potentials
\begin{align}
\mu_{\rm n}= & \frac{\partial\rho}{\partial n_{\rm n}}=\frac{\partial\rho_{\text{nuc}}}{\partial n_{\rm n}}, \\
\mu_{\rm q}= & \frac{\partial\rho}{\partial n_{\rm q}}=\frac{\partial\rho_{\text{nuc}}}{\partial n_{\rm q}}+\frac{\partial\rho_{\rm e}}{\partial n_{\rm e}}\frac{\partial n_{\rm e}}{\partial n_{\rm q}}+\frac{\partial\rho_{\upmu}}{\partial n_{\upmu}}\frac{\partial n_{\upmu}}{\partial n_{\rm q}},
\label{eq:ChargedChemicalPotential}
\end{align}
which are equal in beta equilibrium. 

The motion of the two fluids is described by the relativistic Euler equations
~\citep{Carter1998,Andersson2007} 
\begin{align}
0={}&u_{\rm n}^{\rho}\nabla_{\rho}(\mu_{\rm n}u^n_{\sigma})+\nabla_{\sigma}\mu_{\rm n}-2u_{\rm n}^{\rho}\nabla_{[\rho}(\mu_{\rm n}\epsilon_{\rm n}W_{\sigma]}), \label{eq:EulerN} \\
0={}&u_{\rm q}^{\rho}\nabla_{\rho}(\mu_{\rm q}u_{\sigma}^{\rm q})+\nabla_{\sigma}\mu_{\rm q}+(\mu_{\upmu}-\mu_{\rm e})\nabla_{\sigma}f+2u_{\rm q}^{\rho}\nabla_{[\rho}(\mu_{\rm n}\epsilon_{\rm p}W_{\sigma]}), \label{eq:EulerQ}
\end{align}
and the continuity equations
\begin{align}
\nabla_{\rho}(n_{\rm n}u^{\rho}_{\rm n})=0, \\
\nabla_{\rho}(n_{\rm q}u^{\rho}_{\rm q})=0.
\end{align}
where $W_{\sigma}=u^{\rm n}_{\sigma}-u^{\rm q}_{\sigma}$. $\epsilon_{\rm n}$ and $\epsilon_{\rm p}$ are defined to parameterize the entrainment, and are related by
\begin{equation}
n_{\rm q}\epsilon_{\rm p} = n_{\rm n}\epsilon_{\rm n}.
\end{equation}
The entrainment parameters $\epsilon_{\rm p}$ and $\epsilon_{\rm n}$ here are dimensionless and are the same as those of~\citet{Prix2002}. We vary the parameter $\epsilon_{\rm p}$ to adjust the strength of the entrainment, noting that the effective mass of the proton $m_{\rm p}^*$ is often related to $\epsilon_{\rm p}$ via
\begin{equation}
\epsilon_{\rm p}=1-\frac{m_{\rm p}^*}{m_{\rm N}}.
\end{equation}
The term in Eq.~(\ref{eq:EulerQ}) $\propto \nabla_{\sigma} f$ is responsible for the leptonic buoyancy, and in the outer regions of the core without muons, it is zero.

We now calculate the equations of motion for perturbations to a spherically-symmetric, static background in chemical equilibrium. The metric in Schwarzschild coordinates is
\begin{equation}
{\rm d}s^2=-{\rm e}^{\nu(r)}{\rm d}t^2+{\rm e}^{\lambda(r)}{\rm d}r^2+r^2({\rm d}\theta^2+\sin^2\theta {\rm d}\phi^2),
\end{equation}
where ${\rm e}^{\lambda(r)}=(1-2M(r)/r)^{-1}$, $M(r)$ is the mass enclosed within radius $r$, and ${\rm e}^{\nu(r)}$ is determined using the gravitational redshift formula $\mu(r)\sqrt{-g_{00}}=\text{constant}$, so
\begin{equation}
{\rm e}^{\nu(r)}=(-g_{00})=\left(\frac{m_{N,\text{Fe}-56}}{\mu_0(r)}\right)^2\left(1-\frac{2M}{R}\right),
\end{equation}
where $R$ is the coordinate radius of the star, $M=M(R)$ its total mass computed using the equation of state and the TOV equation and $m_{N,\text{Fe}-56}$ is the mass per baryon of an iron-56 nucleus.

Since the velocities under consideration are much less than the speed of light, we can ignore relativistic gamma factors and write the fluid four-velocities as 
\begin{equation}
u^{\mu}_a=\frac{{\rm d}x^{\mu}_a}{{\rm d}\tau}\approx {\rm e}^{-\nu/2}\frac{{\rm d}x^{\mu}_a}{{\rm d}t}.
\label{eq:FourVelocityDefinition}
\end{equation}
For a stationary background, $u_a^{\mu}={\rm e}^{-\nu/2}(1,0,0,0)$. The velocity of the perturbation $\delta u_a^{\mu}$ to first order in perturbation theory and in the Cowling approximation is thus given by \citep{Andersson2007}
\begin{align}
\delta u_a^{\mu} = {}&(\delta^{\mu}_{\nu}+u_a^{\mu}u^a_{\nu})(u_a^{\sigma}\nabla_{\sigma}\overline{\xi}_a^{\nu}-\overline{\xi}_a^{\sigma}\nabla_{\sigma}u_a^{\nu}) \nonumber \\
={}&(\delta^{\mu}_{\nu}-\delta^{\mu}_0\delta^0_{\nu})\left({\rm e}^{-\nu/2}\nabla_{0}\overline{\xi}_a^{\nu}-\overline{\xi}_a^{\sigma}\nabla_{\sigma}{\rm e}^{-\nu/2}\delta^{\nu}_0\right)
\end{align}
for Lagrangian displacement fields $\overline{\xi}_a^{\mu}$ defined in a coordinate basis. We set $\overline{\xi}_a^0=0$ using the gauge freedom within the definition of $\overline{\xi}^{\mu}_a$. Taking the Eulerian perturbation of Eq.~(\ref{eq:EulerQ}) and considering its spatial components $\sigma=i=1,2,3$, we obtain to first order in perturbation theory
\begin{align}
0={}&{\rm e}^{-\nu}\partial^2_t\overline{\xi}_{\rm q}^i+{\rm e}^{-\nu}\epsilon_{\rm p}\partial^2_t(\overline{\xi}^i_{\rm n}-\overline{\xi}_{\rm q}^i)+g^{ii}\partial_i\left(\frac{\delta\mu_{\rm q}}{\mu_0}\right) \nonumber \\
{}&+\frac{(\delta\mu_{\upmu}-\delta\mu_{\rm e})}{\mu_0}g^{ii}\partial_i f,
\label{eq:PerturbedEulerQ} \\
0={}&{\rm e}^{-\nu}\partial^2_t\overline{\xi}_{\rm n}^i+{\rm e}^{-\nu}\epsilon_{\rm n}\partial^2_t(\overline{\xi}^i_{\rm q}-\overline{\xi}_{\rm n}^i)+g^{ii}\partial_i\left(\frac{\delta\mu_{\rm n}}{\mu_0}\right),
\label{eq:PerturbedEulerN}
\end{align}
where $\mu_0$ is the common background equilibrium chemical potential. The perturbed continuity equations are identical:
\begin{align}
\delta n_a = -n_a\Theta_a-\overline{\xi}^r_a\frac{{\rm d}n_a}{{\rm d}r},
\end{align}
where we have defined
\begin{align}
\Theta_a = \frac{1}{\sqrt{-g}}\frac{\partial(\sqrt{-g}\overline{\xi}_a^i)}{\partial x^i}.
\label{eq:CovariantDivergence}
\end{align}

Since we consider nonrotating stars, spherical symmetry is preserved and the normal modes are spheroidal/poloidal. The displacement field for such a mode in the orthonormal tetrad is
\begin{equation}
\boldsymbol{\xi}_a={\rm e}^{i\omega t}\left[\xi^r_a(r)Y_{lm}(\theta,\phi)\hat{\mathbf{e}}_r+\xi^{\perp}_a(r)r\nabla Y_{lm}(\theta,\phi)\right] \quad a={\rm n,q},
\label{eq:DisplacementField}
\end{equation}
where $Y_{lm}(\theta,\phi)$ are the usual spherical harmonics and we use the usual orthonormal basis vectors. $\omega$ is the angular frequency of the oscillation as observed far from the star. In the coordinate basis, the components of $\overline{\xi}^i_a$ are
\begin{align}
\overline{\xi}^r_a ={}& {\rm e}^{-\lambda/2}\xi^r_a(r)Y_{lm}(\theta,\phi){\rm e}^{i\omega t}, \label{eq:CoordinateBasisXir}\\
\overline{\xi}^{\theta}_a ={}& \xi^{\perp}_a(r)\frac{1}{r}\partial_{\theta}Y_{lm}(\theta,\phi){\rm e}^{i\omega t}, \label{eq:CoordinateBasisXitheta}\\
\overline{\xi}^{\phi}_a ={}& \xi^{\perp}_a(r)\frac{1}{r\sin\theta}\partial_{\phi}Y_{lm}(\theta,\phi){\rm e}^{i\omega t}. \label{eq:CoordinateBasisXiphi}
\end{align}

To compute the buoyant term in Eq.~(\ref{eq:PerturbedEulerQ}), use 
\begin{equation}
\frac{\partial\rho}{\partial f}=\frac{\partial \rho_{\rm e}}{\partial n_{\rm e}}\frac{\partial n_{\rm e}}{\partial f}+\frac{\partial \rho_{\upmu}}{\partial n_{\upmu}}\frac{\partial n_{\upmu}}{\partial f}=n_{\rm q}(\mu_{\rm e}-\mu_{\upmu});
\end{equation}
then
\begin{align}
\delta\mu_{\upmu}-\delta\mu_{\rm e}= & -\delta\left(\frac{1}{n_{\rm q}}\frac{\partial\rho}{\partial f}\right) \nonumber \\ 
= & -\frac{1}{n_{\rm q}}\left(\frac{\partial^2\rho}{\partial f^2}\delta f+\frac{\partial\mu_{\rm q}}{\partial f}\delta n_{\rm q} \right) \nonumber \\
= & \mu_{\rm qf}\Theta_{\rm q}, \label{eq:EulerianPertLeptonChemPotDiff}
\end{align}
where we have defined the thermodynamic derivatives
\begin{equation}
\mu_{ab}\equiv\frac{\partial\mu_a}{\partial n_{\rm b}} \quad a,b\in\{{\rm n,q}\}; \quad\mu_{\rm qf}\equiv\frac{\partial\mu_{\rm q}}{\partial f},\label{eq:MuAB}
\end{equation}
where $\mu_{\rm nq}=\mu_{\rm qn}$; explicitly, 
\begin{align}
\mu_{\rm n} &=\frac{\partial\rho}{\partial n_{\rm b}}-\frac{Y}{n_{\rm b}}\frac{\partial\rho}{\partial Y}, \\
\mu_{\rm q} &=\frac{\partial\rho}{\partial n_{\rm b}}+\frac{(1-Y)}{n_{\rm b}}\frac{\partial\rho}{\partial Y}, \\
\mu_{\rm nn} &=\frac{\partial^2\rho}{\partial n_{\rm b}^2}-\frac{2Y}{n_{\rm b}}\frac{\partial^2\rho}{\partial n_{\rm b}\partial Y}+\frac{Y^2}{n_{\rm b}^2}\frac{\partial^2\rho}{\partial Y^2}, \label{eq:Munn} \\
\mu_{\rm qq} &=\frac{\partial^2\rho}{\partial n_{\rm b}^2}+\frac{2(1-Y)}{n_{\rm b}}\frac{\partial^2\rho}{\partial n_{\rm b}\partial Y}+\frac{(1-Y)^2}{n_{\rm b}^2}\frac{\partial^2\rho}{\partial Y^2}, \label{eq:Muqq}\\
\mu_{\rm nq} &=\frac{\partial^2\rho}{\partial n_{\rm b}^2}+\frac{(1-2Y)}{n_{\rm b}}\frac{\partial^2\rho}{\partial n_{\rm b}\partial Y}-\frac{Y(1-Y)}{n_{\rm b}^2}\frac{\partial^2\rho}{\partial Y^2}. \label{eq:Munq}
\end{align}

Henceforth, we define
\begin{equation}
\Pi_a=\Pi_a(r)Y_{lm}(\theta,\phi)\equiv\frac{\delta\mu_a}{\mu_0},
\label{eq:PiDefinition}
\end{equation}
in terms of which the Euler equations are
\begin{align}
&\omega^2{\rm e}^{-\nu}(1-\epsilon_{\rm n})\xi^r_{\rm n}+\omega^2{\rm e}^{-\nu}\epsilon_{\rm n}\xi^r_{\rm q}={\rm e}^{-\lambda/2}\frac{{\rm d}\Pi_{\rm n}}{{\rm d}r}, \label{eq:XiRN}\\
&\omega^2{\rm e}^{-\nu}r(1-\epsilon_{\rm n})\xi^{\perp}_{\rm n}+\omega^2{\rm e}^{-\nu}r\epsilon_{\rm n}\xi^{\perp}_{\rm q}=\Pi_{\rm n}, \label{eq:XiPerpN} \\
&\omega^2{\rm e}^{-\nu}(1-\epsilon_{\rm p})\xi^r_{\rm q}+\omega^2{\rm e}^{-\nu}\epsilon_{\rm p}\xi^r_{\rm n}={\rm e}^{-\lambda/2}\frac{{\rm d}\Pi_{\rm q}}{{\rm d}r}+{\rm e}^{-\lambda/2}\frac{\mu_{\rm qf}}{\mu_0}\frac{{\rm d}f}{{\rm d}r}\Theta_{\rm q}, \label{eq:XiRQ} \\
&\omega^2{\rm e}^{-\nu}r(1-\epsilon_{\rm p})\xi^{\perp}_{\rm q}+\omega^2{\rm e}^{-\nu}r\epsilon_{\rm p}\xi^{\perp}_{\rm n}=\Pi_{\rm q}. \label{eq:XiPerpQ}
\end{align}
We also recast the continuity equations in terms of $\xi_a^r$ and $\Pi_a$; using
\begin{equation}
\Theta_a = \left(\frac{{\rm e}^{-\lambda/2}}{r^2}\frac{d(r^2\xi^r_a)}{dr}-\frac{l(l+1)}{r}\xi^{\perp}_a\right)Y_{lm},
\end{equation}
we obtain
\begin{align}
\delta n_{\rm n}(r) ={}& -n_{\rm n}\left[\frac{{\rm e}^{-\lambda/2}}{r^2}\frac{{\rm d}(r^2\xi^r_{\rm n})}{{\rm d}r}-{\rm e}^{\nu}\frac{k^2_{\perp}}{\omega^2}\left\{(1+x)\Pi_{\rm n}-x\Pi_{\rm q}\right\}\right] \nonumber \\ 
&-{\rm e}^{-\lambda/2}\xi_{\rm n}^r\frac{{\rm d}n_{\rm n}}{{\rm d}r}, \label{eq:EulerDensPertRBVn} \\
\delta n_{\rm q}(r) ={}& -n_{\rm q}\left[\frac{{\rm e}^{-\lambda/2}}{r^2}\frac{{\rm d}(r^2\xi^r_{\rm q})}{{\rm d}r}-{\rm e}^{\nu}\frac{k^2_{\perp}}{\omega^2}\left\{(1+y)\Pi_{\rm q}-y\Pi_{\rm n}\right\}\right] \nonumber \\ & -{\rm e}^{-\lambda/2}\xi_{\rm q}^r\frac{{\rm d}n_{\rm q}}{{\rm d}r}, \label{eq:EulerDensPertRBVq}
\end{align}
where $k^2_{\perp}\equiv l(l+1)r^{-2}$ and
\begin{align}
x\equiv \frac{\epsilon_{\rm n}}{1-\epsilon_{\rm p}-\epsilon_{\rm n}}, \\
y\equiv \frac{\epsilon_{\rm p}}{1-\epsilon_{\rm p}-\epsilon_{\rm n}}.
\end{align}
Using
\begin{align}
\delta\mu_{\rm n} = {}& \mu_{\rm nn}\delta n_{\rm n}+\mu_{\rm nq}\delta n_{\rm q}, \label{eq:EulerianPertMun} \\
\delta\mu_{\rm q}= {}& \mu_{\rm qq}\delta n_{\rm q}+\mu_{\rm nq}\delta n_{\rm n}+\mu_{\rm qf}\delta f ,\label{eq:EulerianPertMuq}
\end{align}
we find
\begin{align}
\delta n_{\rm n}(r)= & \frac{\mu_{\rm qq}\mu_0\Pi_{\rm n}-\mu_{\rm nq}\mu_0\Pi_{\rm q}-\mu_{\rm nq}\mu_{\rm qf}\xi^r_{\rm q}({\rm d}f/{\rm d}r)}{D}, \label{eq:DensityNEulerianPert} \\
\delta n_{\rm q}(r)= & \frac{\mu_{\rm nn}\mu_0\Pi_{\rm q}-\mu_{\rm nq}\mu_0\Pi_{\rm n}+\mu_{\rm nn}\mu_{\rm qf}\xi^r_{\rm q}({\rm d}f/{\rm d}r)}{D}, \label{eq:DensityQEulerianPert}
\end{align}
where $D\equiv(\mu_{\rm nn}\mu_{\rm qq}-\mu_{\rm nq}^2)$; then Eq.~(\ref{eq:EulerDensPertRBVn}--\ref{eq:EulerDensPertRBVq}) are
\begin{align}
&\frac{{\rm d}\xi^r_{\rm n}}{{\rm d}r}+\left[\frac{2}{r}+\frac{{\rm d}\ln n_{\rm n}}{{\rm d}r}\right]\xi^r_{\rm n}+\left[-\frac{k^2_{\perp}}{\omega^2}{\rm e}^{\nu+\lambda/2}(1+x)+\frac{\mu_0\mu_{\rm qq}}{n_{\rm n}D}{\rm e}^{\lambda/2}\right]\Pi_{\rm n} \nonumber 
\\&=\left[-\frac{k^2_{\perp}}{\omega^2}{\rm e}^{\nu+\lambda/2}x+\frac{\mu_0\mu_{\rm nq}}{n_{\rm n}D}{\rm e}^{\lambda/2}\right]\Pi_{\rm q}+\frac{\mu_{\rm nq}\mu_{\rm qf}}{n_{\rm n}D}\frac{{\rm d}f}{{\rm d}r}\xi^r_{\rm q}, \label{eq:PerturbedContinuityN} \\
&\frac{{\rm d}\xi^r_{\rm q}}{{\rm d}r}+\left[\frac{2}{r}+\frac{{\rm d}\ln n_{\rm q}}{{\rm d}r}+\frac{\mu_{\rm nn}\mu_{\rm qf}}{n_{\rm q}D}\frac{{\rm d}f}{{\rm d}r}\right]\xi^r_{\rm q} \nonumber \\ 
&+\left[-\frac{k^2_{\perp}}{\omega^2}{\rm e}^{\nu+\lambda/2}(1+y)+\frac{\mu_0\mu_{\rm nn}}{n_{\rm q}D}{\rm e}^{\lambda/2}\right]\Pi_{\rm q} \nonumber
\\& =\left[-\frac{k^2_{\perp}}{\omega^2}{\rm e}^{\nu+\lambda/2}y+\frac{\mu_0\mu_{\rm nq}}{n_{\rm q}D}{\rm e}^{\lambda/2}\right]\Pi_{\rm n}. \label{eq:PerturbedContinuityQ}
\end{align}
Notice that, in the case of zero entrainment and zero thermodynamic coupling $\mu_{\rm nq}=0$, the two equations describing the neutron fluid motion (\ref{eq:XiRN}) and~(\ref{eq:XiRQ}) are completely uncoupled from those describing the charged fluid motion (\ref{eq:PerturbedContinuityN}) and~(\ref{eq:PerturbedContinuityQ}), leading to two sets of equations coupling ($\xi^r_{\rm n}$,$\Pi_{\rm n}$) and ($\xi^r_{\rm q}$,$\Pi_{\rm q}$), respectively.

\subsection{Brunt--V\"{a}is\"{a}l\"{a} frequency}
\label{sec:BVFreq}

\begin{figure}
\centering
\includegraphics[width=1.0\columnwidth]{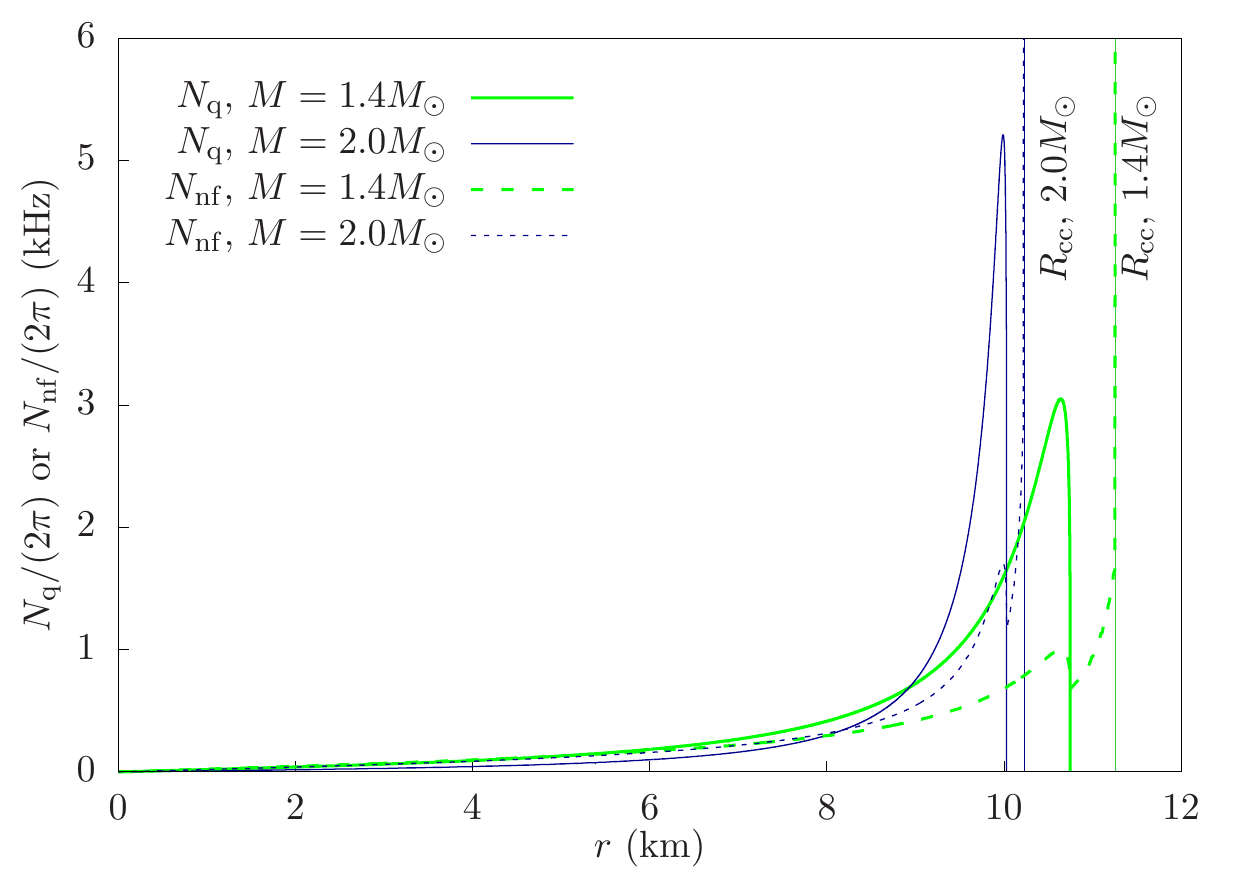}
\caption{Brunt--V\"{a}is\"{a}l\"{a} (cyclical) frequency as a function of coordinate radius $r$ with no entrainment, calculated using the PC1 parametrization for our equation of state from Section~(\ref{sec:EoS}) for two different model stars: $M=1.4M_{\odot}$,  $n_{{\rm b},\text{cntr}}=3.17n_{\text{nuc}}$, $R=12.11$ km, $R_{\text{cc}}=11.27$ km and  $M=2.0M_{\odot}$, $n_{{\rm b},\text{cntr}}=6.67n_{\text{nuc}}$, $R=10.54$ km and $R_{\text{cc}}=10.23$ km, where $n_{{\rm b},\text{cntr}}$ is the central baryon density and $R_{\text{cc}}$ the crust-core interface radius. $R_{\text{cc}}$ for each star is also indicated on the graph by a vertical line at the frequency cutoff for $N_{\text{nf}}$. The Brunt--V\"{a}is\"{a}l\"{a} frequency arising due to the leptonic composition gradient in superfluid stars $N_{\rm q}$ and the total Brunt--V\"{a}is\"{a}l\"{a} frequency for a normal fluid star $N_{\text{nf}}$ are displayed. The frequency cutoffs for $N_{\rm q}$ correspond to the muon threshold, which is at the same density but a different radius for each star.}
\label{fig:BVFrequency}
\end{figure}

\begin{figure}
\centering
\includegraphics[width=1.0\columnwidth]{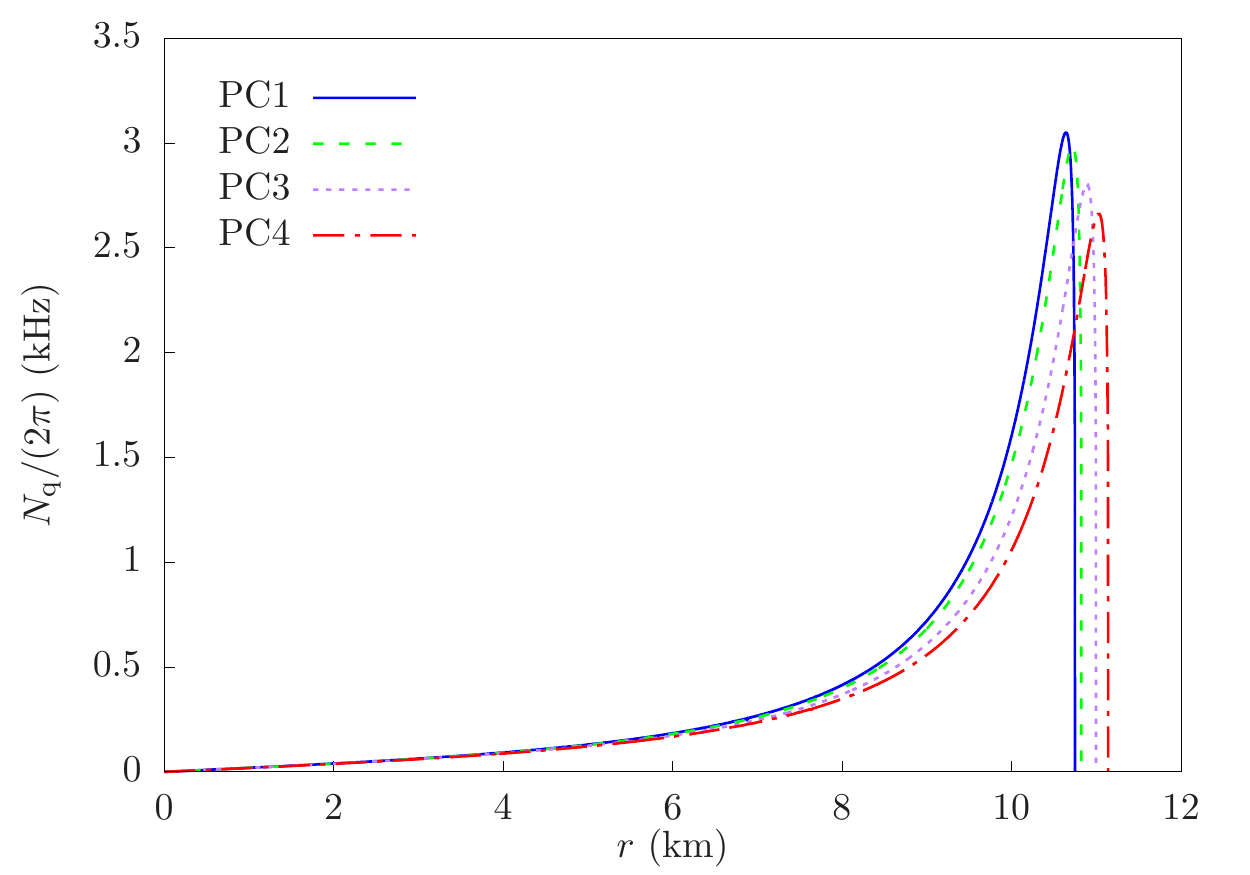}
\caption{Brunt--V\"{a}is\"{a}l\"{a} (cyclical) frequency as a function of coordinate radius $r$ with no entrainment, for fixed mass $1.4M_{\odot}$ using the four EOS parametrizations specified by Table~\ref{tab:EOSParameters}. The frequency cutoffs correspond to the muon threshold, which is at the same density but a different radius for each star.}
\label{fig:BVFrequencyVarK}
\end{figure}

To determine the Brunt--V\"{a}is\"{a}l\"{a} frequency, we use the radial components of the Euler equations to get
\begin{align}
\omega^2{\rm e}^{-\nu}\frac{1}{1+y}\xi^r_{\rm q}={}& {\rm e}^{-\lambda/2}\frac{{\rm d}\Pi_{\rm q}}{{\rm d}r}-{\rm e}^{-\lambda/2}\frac{y}{1+y}\frac{{\rm d}\Pi_{\rm n}}{{\rm d}r} \nonumber \\
& -{\rm e}^{-\lambda/2}\frac{\mu_{\rm qf}(\mu_{\rm nn}\Pi_{\rm q}-\mu_{\rm nq}\Pi_{\rm n})}{n_{\rm q}D}\frac{{\rm d}f}{{\rm d}r} \nonumber \\
&-{\rm e}^{-\lambda}\xi^r_{\rm q}\frac{\mu_{\rm qf}}{\mu_0n_{\rm q}}\frac{{\rm d}f}{{\rm d}r}\left[\frac{{\rm d}n_{\rm q}}{{\rm d}r}+\frac{\mu_{\rm nn}\mu_{\rm qf}}{D}\frac{{\rm d}f}{{\rm d}r}\right];
\label{eq:ChargedFluidRadialComponent2}
\end{align}
Eqs.~(\ref{eq:EulerianPertMun}--\ref{eq:DensityQEulerianPert}) with $\delta\mu_a=(d\mu_a/dr)\delta r$ and $\delta n_a=(dn_a/dr)\delta r$ imply
\begin{equation}
\frac{{\rm d}n_{\rm q}}{{\rm d}r}+\frac{\mu_{\rm nn}\mu_{\rm qf}}{D}\frac{{\rm d}f}{{\rm d}r}=\frac{\mu_{\rm nn}-\mu_{\rm nq}}{D}\frac{{\rm d}\mu_0}{{\rm d}r},
\label{eq:dnqdRReplacement}
\end{equation}
hence
\begin{align}
\xi^r_{\rm q}&\left[\omega^2+{\rm e}^{\nu-\lambda}(1+y)\left(\frac{\mu_{\rm qf}}{\mu_0}\frac{{\rm d}\mu_0}{{\rm d}r}\right)\left(\frac{\mu_{\rm nn}-\mu_{\rm nq}}{n_{\rm q}D}\right)\frac{{\rm d}f}{{\rm d}r}\right] \nonumber \\
={}&{\rm e}^{\nu-\lambda/2}\left((1+y)\frac{{\rm d}\Pi_{\rm q}}{{\rm d}r}-y\frac{{\rm d}\Pi_{\rm n}}{{\rm d}r}\right. \nonumber \\ {}&\left.-(1+y)\frac{\mu_{\rm qf}(\mu_{\rm nn}\Pi_{\rm q}-\mu_{\rm nq}\Pi_{\rm n})}{n_{\rm q}D}\frac{{\rm d}f}{{\rm d}r}\right).
\end{align}
So the square of the Brunt--V\"{a}is\"{a}l\"{a} frequency is
\begin{equation}
N_{\rm q}^2(r)=-{\rm e}^{\nu-\lambda}(1+y)\left(\frac{\mu_{\rm qf}}{\mu_0}\frac{{\rm d}\mu_0}{{\rm d}r}\right)\left(\frac{\mu_{\rm nn}-\mu_{\rm nq}}{n_{\rm q}D}\right)\frac{{\rm d}f}{{\rm d}r}.
\label{eq:BruntVaisalaFrequency1}
\end{equation}
This can be rewritten in a manner which eliminates the dependence on derivatives of $f$. Using $\mu_{\rm e}=\mu_{\upmu}$ in the background, we can write
\begin{align}
(3\pi^2n_{\rm e})^{1/3}={}&\sqrt{(3\pi^2n_{\upmu})^{2/3}+m_{\upmu}^2} \nonumber \\
&\Rightarrow f^{2/3}-(1-f)^{2/3}=\left(\frac{m_{\upmu}^3}{3\pi^2n_{\rm q}}\right)^{2/3}
\label{eq:ElectronFractionRelation}
\end{align}
and
\begin{equation}
\frac{{\rm d}f}{{\rm d}r}=-\frac{f^{1/3}(1-f)^{1/3}}{n_{\rm q}^{5/3}[f^{1/3}+(1-f)^{1/3}]}\left(\frac{m_{\upmu}^2}{(3\pi^2)^{2/3}}\right)\frac{dn_{\rm q}}{dr}.
\label{eq:dfdR}
\end{equation}
Differentiating Eq.~(\ref{eq:ChargedChemicalPotential}) to find
\begin{align}
\mu_{\rm qf}={}&\frac{\partial}{\partial f}\left(\mu_{\rm p}+f\mu_{\rm e}(fn_{\rm q})+(1-f)\mu_{\upmu}((1-f)n_{\rm q})\right) \nonumber \\
={}&n_{\rm e}\frac{{\rm d}\mu_{\rm e}}{{\rm d}n_{\rm e}}-n_{\upmu}\frac{{\rm d}\mu_{\upmu}}{{\rm d}n_{\upmu}}=\frac{m_{\upmu}^2}{3(3\pi^2n_{\rm q}f)^{1/3}},
\label{eq:MuQF2}
\end{align}
we therefore have
\begin{align}
&\mu_{\rm qf}\frac{{\rm d}f}{{\rm d}r}=-\frac{m_{\upmu}}{n_{\rm q}}G(f)\frac{{\rm d}n_{\rm q}}{{\rm d}r}, \nonumber \\
&G(f)\equiv{}\frac{(1-f)^{1/3}}{3}\left(f^{1/3}-(1-f)^{1/3}\right)\left(f^{2/3}-(1-f)^{2/3}\right)^{1/2}.
\label{eq:G(f)}
\end{align}
Inserting this into Eq.~(\ref{eq:dnqdRReplacement}) to eliminate $dn_{\rm q}/dr$, then using the resulting equation to eliminate $\mu_{\rm qf}df/dr$ from Eq.~(\ref{eq:BruntVaisalaFrequency1}) gives us
\begin{equation}
N_{\rm q}^2(r)={\rm e}^{\nu-\lambda}(1+y)\frac{m_{\upmu}G(f)(\mu_{\rm nn}-\mu_{\rm nq})^2({\rm d}\mu_0/{\rm d}r)^2}{\mu_0n_{\rm q}D[n_{\rm q}D-\mu_{\rm nn}m_{\upmu}G(f)]}.
\label{eq:BruntVaisalaFrequency2}
\end{equation}
$N_{\rm q}(r)$ is plotted in Figure~(\ref{fig:BVFrequency}) with zero entrainment ($y=0$), along with the Brunt--V\"{a}is\"{a}l\"{a} frequency for a normal fluid star $N_{\text{nf}}$, given by
\begin{equation}
N_{\text{nf}}=\sqrt{N^2_{Y}+YN_{\rm q}^2},
\label{eq:NFBruntVaisalaFrequency}
\end{equation}
where the lepton gradient contribution is reduced by a factor of $Y$ due to the increased inertia of moving both the protons and neutron, and where~\citep{Reisenegger1992}
\begin{align}
N_{Y}^2(r)={}&-{\rm e}^{\nu-\lambda}\left(\frac{1}{\mu_0}\frac{{\rm d}\mu_0}{{\rm d}r}\right)\left(\frac{\mu_{{\rm b}Y}}{n_{\rm b}\mu_{\rm bb}}\frac{{\rm d}Y}{{\rm d}r}+\frac{Y\mu_{\rm qf}}{n_{\rm b}\mu_{\rm bb}}\frac{{\rm d}f}{{\rm d}r}\right), \label{eq:YGradientBruntVaisalaFrequency} \\
\mu_{\rm bb}={}&Y^2\mu_{\rm qq}+2Y(1-Y)\mu_{\rm nq}+(1-Y)^2\mu_{\rm nn} \nonumber , \\
\mu_{{\rm b}Y}={}&n_{\rm q}(\mu_{\rm qq}-\mu_{\rm nq})-n_{\rm n}(\mu_{\rm nn}-\mu_{\rm nq}). \nonumber
\end{align}
The superfluid leptonic Brunt--V\"{a}is\"{a}l\"{a} frequency is similar to that of Figure 2 of KG14, Figure 5 of~\citet{Passamonti2016} and the zero entrainment results of YW17. Differences are due to differences in the equations of state used, and in the case of the YW17 results, our inclusion of general relativity in both the background star and the perturbations and our neglect of self-gravity perturbations. Figure~\ref{fig:BVFrequencyVarK} shows $N_{\rm q}$ for fixed mass $M=1.4M_{\odot}$ using each of the four parametrizations specified by Table~\ref{tab:EOSParameters}, with the peak value of $N_{\rm q}$ decreasing slightly with increasing nuclear compressibility $K$.

Using the definition of the Brunt--V\"{a}is\"{a}l\"{a} frequency, we can rewrite Eq.~(\ref{eq:XiRQ}) as
\begin{align}
\xi^r_{\rm q} {\rm e}^{-\nu}(\omega^2-N_{\rm q}^2)={}&{\rm e}^{-\lambda/2}(1+y)\frac{{\rm d}\Pi_{\rm q}}{{\rm d}r}-{\rm e}^{-\lambda/2}y\frac{{\rm d}\Pi_{\rm n}}{{\rm d}r} \nonumber \\
&+\frac{{\rm e}^{\lambda/2-\nu}\mu_0N_{\rm q}^2(\mu_{\rm nn}\Pi_{\rm q}-\mu_{\rm nq}\Pi_{\rm n})}{({\rm d}\mu_0/{\rm d}r)(\mu_{\rm nn}-\mu_{\rm nq})}.
\label{eq:XiRQ2}
\end{align}
Eqs.~(\ref{eq:XiRN}),~(\ref{eq:PerturbedContinuityN}--\ref{eq:PerturbedContinuityQ}) and~(\ref{eq:XiRQ2}) are the four coupled first-order ODEs describing the fluid perturbations in the core.

\subsection{Two-fluid formalism in the crust}
\label{subsec:CrustEoM}

To correctly calculate the compressional modes, we need to allow the oscillations in the core to propagate into the crust. As in the core, we consider two fluids- superfluid free neutrons, with displacement field $\overline{\xi}^i_{\rm f}$, and normal fluid nuclei, with displacement field $\overline{\xi}^i_{\rm c}$. We ignore elastic stresses here for simplicity, as we are not interested in the $s$ (shear) and $i$ (interface) modes caused by elasticity in the crust \citep{McDermott1983}. We also neglect entrainment, which is expected in the crust~\citep{Kobyakov2013,Chamel2017a} but which we do not expect to play a large role in the $g$-modes. Its effect on the $p$-modes, which we do not expect to be large either, will be examined in a later paper.

We derive the two-fluid crust equations of motion by first assuming that the perturbations of the chemical potential $\mu_{\rm f}$ and $\mu_{\rm c}$ in the crust can be written as a function of the two crustal number densities $n_{\rm f}$ (free neutrons) and $n_{\rm c}$ (baryons in nuclei), with changes in the other parameters of the crust EOS $A$, $Y$ and $n_{\rm i}$ having been absorbed into the changes in either $n_{\rm f}$ or $n_{\rm c}$. Thus we can write a series of four equations-- the two radial and two tangential components of the perturbed Euler equations-- analogously to Eqs.~(\ref{eq:XiRN}--\ref{eq:XiPerpQ}). We then have
\begin{align}
&\omega^2{\rm e}^{-\nu}\xi^r_{\rm f}={\rm e}^{-\lambda/2}\frac{{\rm d}\Pi_{\rm f}}{{\rm d}r}, \label{eq:XiRF}\\
&\omega^2{\rm e}^{-\nu}r\xi^{\perp}_{\rm f}=\Pi_{\rm f}, \label{eq:XiPerpF} \\
&\omega^2{\rm e}^{-\nu}\xi^r_{\rm c}={\rm e}^{-\lambda/2}\frac{{\rm d}\Pi_{\rm c}}{{\rm d}r}, \label{eq:XiRC} \\
&\omega^2{\rm e}^{-\nu}r\xi^{\perp}_{\rm c}=\Pi_{\rm c}, \label{eq:XiPerpC}
\end{align}
where $\Pi_{\rm f}\equiv\delta\mu_{\rm f}/\mu_0$, $\Pi_{\rm c}\equiv\delta\mu_{\rm c}/\mu_0$ in analogy with Eq.~(\ref{eq:PiDefinition}). From the perturbed continuity equation, we also have
\begin{align}
\delta n_{\rm f} = -n_{\rm f}\Theta_{\rm f}-{\rm e}^{-\lambda/2}\xi^r_{\rm f}\frac{{\rm d}n_{\rm f}}{{\rm d}r}, \label{eq:EulerDensPertnf}\\
\delta n_{\rm c} = -n_{\rm c}\Theta_{\rm c}-{\rm e}^{-\lambda/2}\xi^r_{\rm c}\frac{{\rm d}n_{\rm c}}{{\rm d}r}. \label{eq:EulerDensPertnc}
\end{align}
Using our assumption about the number density dependence of the perturbations, we can then rearrange
\begin{align}
\delta\mu_{\rm f}={}& \mu_{\rm ff}\delta n_{\rm f}+\mu_{\rm fc}\delta n_{\rm c}, \\
\delta\mu_{\rm c}={}& \mu_{\rm cc}\delta n_{\rm c}+\mu_{\rm fc}\delta n_{\rm f}
\end{align}
to obtain
\begin{align}
\delta n_{\rm f}=\frac{\mu_0(\mu_{\rm cc}\Pi_{\rm f}-\mu_{\rm fc}\Pi_{\rm c})}{D_{\text{crust}}}, \label{eq:deltanf}\\
\delta n_{\rm c}=\frac{\mu_0(\mu_{\rm ff}\Pi_{\rm c}-\mu_{\rm fc}\Pi_{\rm f})}{D_{\text{crust}}}, \label{eq:deltanc}
\end{align}
where $\mu_{ab}\equiv\partial \mu_a/\partial n_{\rm b}$ for $a,b\in\{\rm f,c\}$, $D_{\text{crust}}\equiv \mu_{\rm ff}\mu_{\rm cc}-\mu_{\rm fc}^2$ and $\mu_0=\mu_{\rm f}=\mu_{\rm c}$ in the background. There are subtleties involved in the calculation of $\mu_{\rm ff}$, $\mu_{\rm cc}$ and $\mu_{\rm fc}$, which are discussed in Appendix~\ref{app:1}. Combining Eqs.~(\ref{eq:deltanf}--\ref{eq:deltanc}) with Eq.~(\ref{eq:EulerDensPertnf}--\ref{eq:EulerDensPertnc}), we obtain
\begin{align}
\frac{{\rm d}\xi^r_{\rm f}}{{\rm d}r}+\left[\frac{2}{r}+\frac{{\rm d}\ln n_{\rm f}}{{\rm d}r}\right]\xi^r_{\rm f}+\left[-{\rm e}^{\nu+\lambda/2}\frac{k^2_{\perp}}{\omega^2}+\frac{\mu_0\mu_{\rm cc}}{n_{\rm f}D_{\text{crust}}}{\rm e}^{\lambda/2}\right]\Pi_{\rm f} \nonumber \\
= \frac{\mu_0\mu_{\rm fc}}{n_{\rm f}D_{\text{crust}}}{\rm e}^{\lambda/2}\Pi_{\rm c}, \label{eq:PerturbedContinuityF}\\
\frac{{\rm d}\xi^r_{\rm c}}{{\rm d}r}+\left[\frac{2}{r}+\frac{{\rm d}\ln n_{\rm c}}{{\rm d}r}\right]\xi^r_{\rm c}+\left[-{\rm e}^{\nu+\lambda/2}\frac{k^2_{\perp}}{\omega^2}+\frac{\mu_0\mu_{\rm ff}}{n_{\rm c}D_{\text{crust}}}{\rm e}^{\lambda/2}\right]\Pi_{\rm c} \nonumber \\
= \frac{\mu_0\mu_{\rm fc}}{n_{\rm c}D_{\text{crust}}}{\rm e}^{\lambda/2}\Pi_{\rm f}. \label{eq:PerturbedContinuityC}
\end{align}
Eqs.~(\ref{eq:XiRF}),~(\ref{eq:XiRC}) and (\ref{eq:PerturbedContinuityF}--\ref{eq:PerturbedContinuityC}) are the four coupled first-order ODEs describing the fluid perturbations in the crust. Similarly to the equations in the core, in the case of zero thermodynamic coupling $\mu_{\rm fc}=0$, these equations become two sets of coupled equations for ($\xi^r_{\rm f}$,$\Pi_{\rm f}$) and ($\xi^r_{\rm c}$,$\Pi_{\rm c}$).

\subsection{Single normal fluid in crust}
\label{subsec:SFCrustEoM}

In the crust, the neutron superfluid is $_1S^0$, with gaps $\sim 0.1$--$1$ MeV until low free neutron density~\citep{Gezerlis2010,Gezerlis2014}. Thus, we expect the free neutron gas to remain superfluid throughout most of the crust. The superfluid gap for neutrons falls precipitously at a Fermi wave number of approximately $0.05$ fm$^{-1}$ \citep{Gezerlis2014}, corresponding to a free neutron number density of $n_{\rm f}=n_{\rm NF}=2.64\times10^{-5}n_{\text{nuc}}$. For our calculations, we assume that the critical temperature $T_{\rm c}$ for free crustal neutrons is purely a function of $n_{\rm f}$. Thus, we assume that the free crustal neutrons are superfluid at $n_{\rm f}>n_{\rm NF}$ and normal at $n_{\rm f}\leq n_{\rm NF}$, so there is a sharp    transition from superfluid to normal at $n_{\rm f}=n_{\rm NF}$. More realistically, $T_{\rm c}$ will also depend on $n_{\rm c}$; we assume that $T_{\rm c}$ depends on $n_{\rm c}$ much more weakly than it does on $n_{\rm f}$. In a finite temperature star, the superfluid density is proportional to $T_{\rm c}-T$ near the transition to normal fluid, but this temperature dependence is only important in a very thin region in the crust for $T$ small compared with typical values of $T_{\rm c}$ in the crust, which are $\sim 10^{10}$ K.

Between the transition density and neutron drip, free neutrons and nuclei move together as a single fluid. We represent this single normal fluid, which exists only in a very thin layer just above the neutron drip density, using the displacement field $\overline{\xi}^i_{\rm b}$ and the total baryon number density $n_{\rm b}=n_{\rm c}+n_{\rm f}$. The equation of state in this region is the same as in the two-fluid region of the crust. Since the two fluids move together here, there is a buoyancy and Brunt--V\"{a}is\"{a}l\"{a} frequency associated with the gradient of $Y_{\rm c}\equiv n_{\rm c}/n_{\rm b}$.


By analogy with Eqs.~(\ref{eq:XiRN}--\ref{eq:XiPerpQ}), ~(\ref{eq:PerturbedContinuityN}--\ref{eq:PerturbedContinuityQ}) and~(\ref{eq:XiRQ2}), we obtain two Euler equations and the perturbed continuity equation for the single normal fluid displacement field in the crust:
\begin{align}
&\omega^2{\rm e}^{-\nu}r\xi^{\perp}_{\rm b} = \Pi_{\rm b}, \label{eq:XiPerpB} \\
&\xi^r_{\rm b}{\rm e}^{-\nu}(\omega^2-N_{\rm b}^2) = {\rm e}^{-\lambda/2}\frac{{\rm d}\Pi_{\rm b}}{{\rm d}r}+\frac{{\rm e}^{\lambda/2-\nu}\mu_0n_{\rm b}^2}{{\rm d}\mu_0/{\rm d}r}\Pi_{\rm b}, \label{eq:XiRB} \\
&\frac{{\rm d}\xi^r_{\rm b}}{{\rm d}r}+\left[\frac{2}{r}+\frac{{\rm d}\ln n_{\rm b}}{{\rm d}r}+\frac{\mu_{bY_{\rm c}}}{n_{\rm b}\mu_{\rm bb}}\frac{{\rm d}Y_{\rm c}}{{\rm d}r}\right]\xi^r_{\rm b} \nonumber \\
&+\left[-{\rm e}^{\nu+\lambda/2}\frac{k_{\perp}^2}{\omega^2}+\frac{\mu_0}{n_{\rm b}\mu_{\rm bb}}{\rm e}^{\lambda/2}\right]\Pi_{\rm b}=0, \label{eq:PerturbedContinuityB}
\end{align}
where $\Pi_{\rm b}\equiv\delta\mu_{\rm b}/\mu_0$ and $n_{\rm b}$ is the Brunt--V\"{a}is\"{a}l\"{a} frequency associated with the gradient of $Y_{\rm c}$, given by
\begin{equation}
N_{\rm b}^2 = -{\rm e}^{\nu-\lambda}\frac{1}{\mu_0}\frac{{\rm d}\mu_0}{{\rm d}r}\frac{\mu_{{\rm b}Y_{\rm c}}}{n_{\rm b}\mu_{\rm bb}}\frac{{\rm d}Y_{\rm c}}{{\rm d}r}.
\label{eq:CrustBVFrequency}
\end{equation}
The two thermodynamic derivatives $\mu_{\rm bb}$ and $\mu_{{\rm b}Y_{\rm c}}$ are
\begin{align}
\mu_{\rm bb} ={}& Y_{\rm c}^2\mu_{\rm cc}+(1-Y_{\rm c})^2\mu_{\rm ff}+2Y_{\rm c}(1-Y_{\rm c})\mu_{\rm fc},\\
\mu_{{\rm b}Y_{\rm c}} ={}& n_{\rm c}(\mu_{\rm cc}-\mu_{\rm fc})-n_{\rm f}(\mu_{\rm ff}-\mu_{\rm fc}).
\end{align}

\subsection{Interface and boundary conditions}

\label{sec:BCs}

At the centre of the star, we impose the regularity condition $\Theta_a=0$, which implies that the displacement fields and $\Pi_a$ satisfy the following conditions at $r=0$:
\begin{align}
\xi^r_a ={}& l(\xi^r_a)_0r^{l-1}, \\
\Pi_a ={}& \omega^2{\rm e}^{-\nu}(\xi^r_a)_0r^l,
\end{align}
where $(\xi^r_a)_0$ is a constant. Since we can scale the overall amplitude of each mode, we only need to specify $(\xi^r_{\rm n})_0$ and can set $(\xi^r_{\rm q})_0=1$.

We require four conditions at the crust-core transition which allow the computation of the four quantities $(\xi^r_{\rm c},\xi^r_{\rm f},\Pi_{\rm c},\Pi_{\rm f})$ on the crust side of the transition using the quantities $(\xi^r_{\rm q},\xi^r_{\rm n},\Pi_{\rm q},\Pi_{\rm n})$ on the core side of the transition. Since the crust-core interface is denoted by the formation of nuclei, we know that the radial component of the displacement fields for the protons must be continuous at this interface. Since the motion of the protons is described by $\xi_{\rm q}^i$ and $\xi_{\rm c}^i$, this implies
\begin{equation}
(\xi^r_{\rm q})^+ = (\xi^r_{\rm c})^-,
\label{eq:InterfaceCondition1}
\end{equation}
where $+$ indicates the high-density (core) side and $-$ the low-density (crust) side of the transition. As baryons are not allowed to build up at the interface, baryon conservation is the second transition condition. Denoting the Lagrangian perturbation moving along with the nuclei (and hence the crust-core boundary) as $\Delta_{\rm c}$, the perturbed continuity equation for the total baryon number density is
\begin{equation}
\Delta_{\rm c}n_{\rm b}+n_{\rm b}\Theta_{\rm c}=0.
\end{equation}
Integrating this across the crust-core interface, we obtain
\begin{equation}
(n_{\rm n}\xi^r_{\rm n}-n_{\rm n}\xi^r_{\rm q})^+=(n_{\rm f}\xi^r_{\rm f}-n_{\rm f}\xi^r_{\rm c})^-.
\label{eq:InterfaceCondition2}
\end{equation}
As we have neglected elastic stresses in the crust, continuity of the tractions across the crust-core interface is given by the continuity of the pressure perturbation moving with the interface, or
\begin{equation}
(\Delta_{\rm c}P)^+=(\Delta_{\rm c}P)^-.
\end{equation}
Using the Gibbs--Duhem equation, we can use $\Delta P = \sum_{a}n_a\Delta \mu_a$ to rewrite this condition, giving
\begin{equation}
(n_{\rm n}\Pi_{\rm n}+n_{\rm q}\Pi_{\rm q})^+=(n_{\rm c}\Pi_{\rm c}+n_{\rm f}\Pi_{\rm f})^-+(n_{\rm b}^--n_{\rm b}^+){\rm e}^{-\lambda/2}\xi^r_{\rm c}\frac{{\rm d}\ln\mu_0}{{\rm d}r}.
\label{eq:InterfaceCondition3}
\end{equation}
Following~\citet{Andersson2011} and~\citet{Passamonti2012}, the final boundary condition we impose is continuity of the neutron chemical potential perturbation, $(\Delta_{\rm c}\mu_{\rm n})^+=(\Delta_{\rm c} \mu_{\rm f})^-$, which results from the ``chemical gauge''-independence of the neutron chemical potential. This final interface condition simplifies to
\begin{equation}
(\Pi_{\rm n})^+ = (\Pi_{\rm f})^-,
\label{eq:InterfaceCondition4}
\end{equation}
where we have used $\mu_0=\mu_{\rm c}=\mu_{\rm f}$ in the background equilibrium. The chemical potential is the same for crustal neutrons that are bound in nuclei or in the surrounding free superfluid; this condition is satisfied in the crustal equation of state, which allows neutrons to be exchanged freely between nuclei and the surrounding free neutron vapor (see Section~\ref{subsec:CrustEoS} and references therein). Thus, Eq.~(\ref{eq:InterfaceCondition4}) states the condition that there is no energy change when a crustal neutron is exchanged with a core neutron at the crust-core boundary irrespective of whether the crustal neutron is bound or free.


At the two fluid-single fluid transition in the crust just above neutron drip, baryon conservation and continuity of the tractions must be imposed. These two conditions can be expressed as
\begin{align}
(n_{\rm f}\xi^r_{\rm f}+n_{\rm c}\xi^r_{\rm c})^+={}&(n_{\rm b}\xi^r_{\rm b})^-,
\label{eq:TFSFCondition1}\\
(n_{\rm f}\Pi_{\rm f}+n_{\rm c}\Pi_{\rm c})^+={}&(n_{\rm b}\Pi_{\rm b})^-,\label{eq:TFSFCondition2}
\end{align}
where $+/-$ indicate the high density (two fluid) and low density (single fluid) regions respectively.
We also require another boundary condition at the two fluid-single fluid transition (SFT). In a very thin region where the superfluid neutron fraction $f$ falls from one to zero, $\xi_{\rm f}^r=f\xi_{\rm sf}^r+(1-f)\xi_{\rm nf}^r$, where $\xi_{\rm sf}^r$ and $\xi_{\rm nf}^r$ are the radial components of the superfluid neutron and normal fluid neutron displacement fields. If the normal neutrons couple perfectly to the charged component, then $f\xi_{\rm sf}^r=\xi_{\rm f}^r-(1-f)\xi_{\rm c}^r\to\xi_{\rm f}^r-\xi_{\rm c}^r$ for $f\to 0$. The current carried by the superfluid component should vanish where the superfluid neutrons disappear, which is true if
\begin{equation}
(\xi^r_{\rm f})^+=(\xi^r_{\rm c})^+
\label{eq:XifXicequal}
\end{equation}
at the surface where $f=0$. Combined with Eq.~(\ref{eq:TFSFCondition1}), this implies that 
\begin{equation}
(\xi^r_{\rm f})^+=(\xi^r_{\rm c})^+=(\xi^r_{\rm b})^-.
\label{eq:ThreeComovingXis}
\end{equation} 

A boundary condition is required at the outer surface of the star, which we approximate to occur at the neutron drip line since the outer crust contains so little of the star's mass (less than $0.01$\%) that we assume that its effect on the modes is negligible. We impose a form of the condition expressed in Eqs.~(\ref{eq:InterfaceCondition3}), but applied to the displacement field of the single normal fluid which exists just above neutron drip (ND)
\begin{align}
\left(\Pi_{\rm b} + {\rm e}^{-\lambda/2}\xi^r_{\rm b}\frac{{\rm d}\ln\mu_0}{{\rm d}r}\right)_{\text{at ND}}=0.
\label{eq:DeltaP}
\end{align}

As a check, we also compute a few $g$-modes and $p$-modes while integrating out to lower densities in the crust, imposing Eq.~(\ref{eq:DeltaP}) at $n_{\rm b}/n_{\text{nuc}}=1\times10^{-8}$. The $g$-mode frequencies obtained when doing so agree to within 0.1\% of those found when we stopped the integration at neutron drip. There is no discernible change in the core displacement fields for $g$-modes for these two boundary conditions, and the changes in the displacement fields in the crust are larger than in the core but still very small. The $p$-mode frequencies obtained in this way are within 2\% of those calculated with neutron drip as the stopping point for the integration. The $p$-mode displacement fields in the core are weakly affected by this shift in the minimum density, but the modes in the crust can differ significantly, particularly for the higher frequency modes which can have additional oscillations in the crust.

\section{Normal mode calculations}
\label{sec:NormalModes}

\subsection{WKB solutions}

Since the leptonic Brunt--V\"{a}is\"{a}l\"{a} frequency does not exist in the crust, we expect that the $g$-mode displacement fields in the crust will be evanescent and nearly zero. We thus employed the WKB approximation to calculate approximate $g$-mode displacement fields and mode frequencies, assuming no propagation into the crust, and also use the resulting approximate $p$-mode dispersion relations in discussing the $p$-mode displacement fields in the core. First, we convert Eqs.~(\ref{eq:XiRN},\ref{eq:PerturbedContinuityN}--\ref{eq:PerturbedContinuityQ},\ref{eq:XiRQ2}) into two second-order equations for $\Pi_{\rm n}$ and $\Pi_{\rm q}$, neglecting curvature terms, derivatives of the metric, $f$, the $\mu_{ab}$ and $N_{\rm q}$, and ignoring entrainment. We obtain
\begin{align}
&\frac{{\rm d}^2\Pi_{\rm n}}{{\rm d}r^2}+\frac{{\rm d}\ln n_{\rm n}}{{\rm d}r}\frac{{\rm d}\Pi_{\rm n}}{{\rm d}r}+\left[-k_{\perp}^2{\rm e}^{\lambda}+\frac{\mu_0\mu_{\rm qq}{\rm e}^{\lambda-\nu}\omega^2}{n_{\rm n}D}\right]\Pi_{\rm n} 
\nonumber \\
{}&= \frac{\mu_{\rm nq}\mu_0{\rm e}^{\lambda-\nu}\omega^2}{n_{\rm n}D}\Pi_{\rm q},
\label{eq:2ndOrderN1} \\
&\frac{{\rm d}^2\Pi_{\rm q}}{{\rm d}r^2}+\frac{{\rm d}\ln n_{\rm q}}{{\rm d}r}\frac{{\rm d}\Pi_{\rm q}}{{\rm d}r}+\left(1-\frac{N_{\rm q}^2}{\omega^2}\right)\left[-k_{\perp}^2{\rm e}^{\lambda} +\frac{\mu_0\mu_{\rm nn}{\rm e}^{\lambda-\nu}\omega^2}{n_{\rm q}D}\right]\Pi_{\rm q} \nonumber \\
 {}&= \frac{\mu_{\rm nq}\mu_0{\rm e}^{\lambda-\nu}(\omega^2-N_{\rm q}^2)}{n_{\rm q}D}\Pi_{\rm n}, \label{eq:2ndOrderQ1}
\end{align}
where we have used Eq.~(\ref{eq:BruntVaisalaFrequency1}) to replace ${\rm d}\mu_0/{\rm d}r$.
Defining $\Psi_a=\sqrt{n_a}\Pi_a$ and assuming that the $\Psi_a$ have a slowly-varying amplitude $C_a(r)$ and a rapidly-oscillating phase $S(r)=\int k_rdr$. Inserting this definition into Eqs.~(\ref{eq:2ndOrderN1}--\ref{eq:2ndOrderQ1}) gives
\begin{align}
(S')^2C_{\rm n} = M_{\rm nn}C_{\rm n} + M_{\rm nq}C_{\rm q}, \label{eq:WKB0O_1}\\
(S')^2C_{\rm q} = M_{\rm qq}C_{\rm q} + M_{\rm qn}C_{\rm n}, \label{eq:WKB0O_3}
\end{align}
where $d/dr='$ and
\begin{align}
M_{\rm nn} &= -k^2_{\perp}{\rm e}^{\lambda}+\frac{{\rm e}^{\lambda-\nu}\omega^2\mu_0\mu_{\rm qq}}{n_{\rm n}D}-\frac{1}{\sqrt{n_{\rm n}}}\frac{{\rm d}^2\sqrt{n_{\rm n}}}{{\rm d}r^2}, \\
M_{\rm qq} &= -\left(1-\frac{N_{\rm q}^2}{\omega^2}\right)k^2_{\perp}{\rm e}^{\lambda}+\frac{{\rm e}^{\lambda-\nu}\omega^2\mu_0\mu_{\rm nn}}{n_{\rm q}D}-\frac{1}{\sqrt{n_{\rm q}}}\frac{{\rm d}^2\sqrt{n_{\rm q}}}{{\rm d}r^2}, \\
M_{\rm nq} &= M_{qn} = -\frac{{\rm e}^{\lambda-\nu}\omega^2\mu_0\mu_{\rm nq}}{\sqrt{n_{\rm n}n_{\rm q}}D}.
\end{align}
Eqs.~(\ref{eq:WKB0O_1}--\ref{eq:WKB0O_3}) have solutions
\begin{equation}
(S')^2 =\frac{1}{2}\left[(M_{\rm nn}+M_{\rm qq}) \pm \sqrt{(M_{\rm nn}-M_{\rm qq})^2+4M_{\rm nq}^2} \right].
\label{eq:ZerothOrderPhase}
\end{equation}
As $|M_{\rm nq}| \ll |M_{\rm nn}|, |M_{\rm qq}|$, we can identify a neutron-dominated mode with $(S')^2=(k_r^2)_+\approx M_{\rm nn}$ and a charged fluid-dominated mode with $(S')^2=(k_r^2)_-\approx M_{\rm qq}$. In the low frequency $\omega^2\lesssim N^2_{\rm q}$ limit, $M_{\rm nn}\sim-k_{\perp}^2{\rm e}^{\lambda}$, so the low-frequency neutron-dominated mode is nonpropagating. The charged fluid-dominated mode does propagate, however, since $M_{\rm qq}\sim (N_{\rm q}^2/ \omega^2-1)k_{\perp}^2 {\rm e}^{\lambda}$ in the low-frequency limit, thus giving a dispersion relation for the $g$-modes
\begin{equation}
\omega^2_g\approx\frac{N_{\rm q}^2k_{\perp}^2{\rm e}^{\lambda}}{k^2},
\label{eq:ApproximateGModeDispersion}
\end{equation}
where $k^2=k_r^2+k^2_{\perp}{\rm e}^{\lambda}$, in agreement with the standard result of~\citet{McDermott1983}. The high frequency limit $\omega^2\gg N_{\rm q}^2$, keeping the $M_{\rm nq}$ contribution, gives the $p$-mode dispersion relation in the crust
\begin{align}
{}& \omega^2_{\rm p}\approx c_{\rm s\pm}^2k^2, \nonumber \\
{}&c_{\rm s\pm}^2={\rm e}^{\nu-\lambda}\frac{n_{\rm n}n_{\rm q}}{2\mu_0}\left[\left(\frac{\mu_{\rm qq}}{n_{\rm n}}+\frac{\mu_{\rm nn}}{n_{\rm q}}\right)\pm\sqrt{\left(\frac{\mu_{\rm qq}}{n_{\rm n}}-\frac{\mu_{\rm nn}}{n_{\rm q}}\right)^2+\frac{4\mu_{\rm nq}^2}{n_{\rm n}n_{\rm q}}}\right],
\label{eq:ApproximatePModeDispersion}
\end{align}
suggesting two sets of $p$-modes, one associated with each superfluid. This is similar to the simplified $p$-mode dispersion relation given by~\citet{Passamonti2016}. Here we have implicitly assumed that the phases of the two fluids are the same. If the thermodynamic coupling $\mu_{\rm nq}$ is ignored, Eq.~(\ref{eq:ApproximatePModeDispersion}) gives two completely separate dispersions, one for the charged fluid and one for the neutron fluid.

In the inner region of the star $r<r_t$, $k_r^2<0$ and the normal modes are exponentially damped. In the outer region $r_t<r<r_{\text{out}}$,  $k_r^2>0$ and the modes are oscillatory. Matching at $r_t$ with the exponential solution in the inner region and imposing $\Psi_a(r_{\text{out}})=0$ assuming no propagation into the crust, allowed $g$-modes will have $k_r$ satisfying
\begin{equation}
\int_{r_t}^{r_{\text{out}}}k_r(r')dr'=\left(n_r-\frac{1}{4}\right)\pi, \quad n_r=1,2,3,...,
\label{eq:krCondition}
\end{equation}
where $n_r$ is the radial node number. This condition determines the allowed frequencies since $k_r(r)$ is a function of $\omega$. $n_r$ here is the radial index of the solution, setting the radial node number for $\Psi_a$ and by extension $\Pi_a$ and $\xi^r_a$.

\subsection{Numerical results}
\subsubsection{$g$-modes}

To obtain solutions for the displacement fields $\xi^i_a$ and the $\Pi_a$, we numerically integrated the system of four first-order equations given in the core by Eqs.~(\ref{eq:XiRN}),~(\ref{eq:PerturbedContinuityN}--\ref{eq:PerturbedContinuityQ}) and~(\ref{eq:XiRQ2}), in the crust by Eqs.~(\ref{eq:XiRF}),~(\ref{eq:XiRC}) and~(\ref{eq:PerturbedContinuityF}--\ref{eq:PerturbedContinuityC}), and in the crust just above neutron drip by Eqs.~(\ref{eq:XiRB}--\ref{eq:PerturbedContinuityB}). We use a standard energy normalization to set the amplitude of the displacement fields. Reinserting factors of $c$, this condition is
\begin{align}
\frac{\omega^2}{c^2}\int_0^{R_{\text{cc}}}\int_{\Omega} {\rm d}V\mu_0\left[ (1-\epsilon_{\rm p})n_{\rm q}\xi^{*i}_{\rm q}\xi^{\rm q}_i +(1-\epsilon_{\rm n})n_{\rm n}\xi^{*i}_{\rm n}\xi^{\rm n}_{\rm i} \right. \nonumber \\ 
\left. + n_{\rm q}\epsilon_{\rm p}(\xi^{*i}_{\rm q}\xi^{\rm n}_{\rm i}+\xi^{*i}_{\rm n}\xi^{\rm q}_i)\right] \nonumber \\
+\frac{\omega^2}{c^2}\int_{R_{\text{cc}}}^R\int_{\Omega} {\rm d}V\mu_0(n_{\rm f}\xi^{*i}_{\rm f}\xi^{\rm f}_i+n_{\rm c}\xi^{*i}_{\rm c}\xi^{\rm c}_i) =\frac{GM^2}{R},
\end{align}
where $M$ and $R$ are the mass and radius of the star, $R_{\text{cc}}$ is the coordinate radius of the crust-core transition, $\Omega$ indicates integration over the solid angle of a sphere and ${\rm d}V={\rm e}^{\lambda/2}r^2\sin\theta {\rm d}r{\rm d}\theta {\rm d}\phi$. In the very thin single fluid region at densities just above neutron drip, $\xi^i_{\rm f}=\xi^i_{\rm c}=\xi^i_{\rm b}$. Each function ($\xi^r_{\rm n}$,$\xi^r_{\rm q}$,$\xi^r_{\rm c}$,$\xi^r_{\rm f}$,$\Pi_{\rm n}$,$\Pi_{\rm q}$,$\Pi_{\rm c}$,$\Pi_{\rm f}$,$\xi^r_{\rm b}$,$\Pi_{\rm n}$) is scaled by the same amount, since they are all linearly related.

\begin{figure*}
\centering
\includegraphics[width=1.0\textwidth]{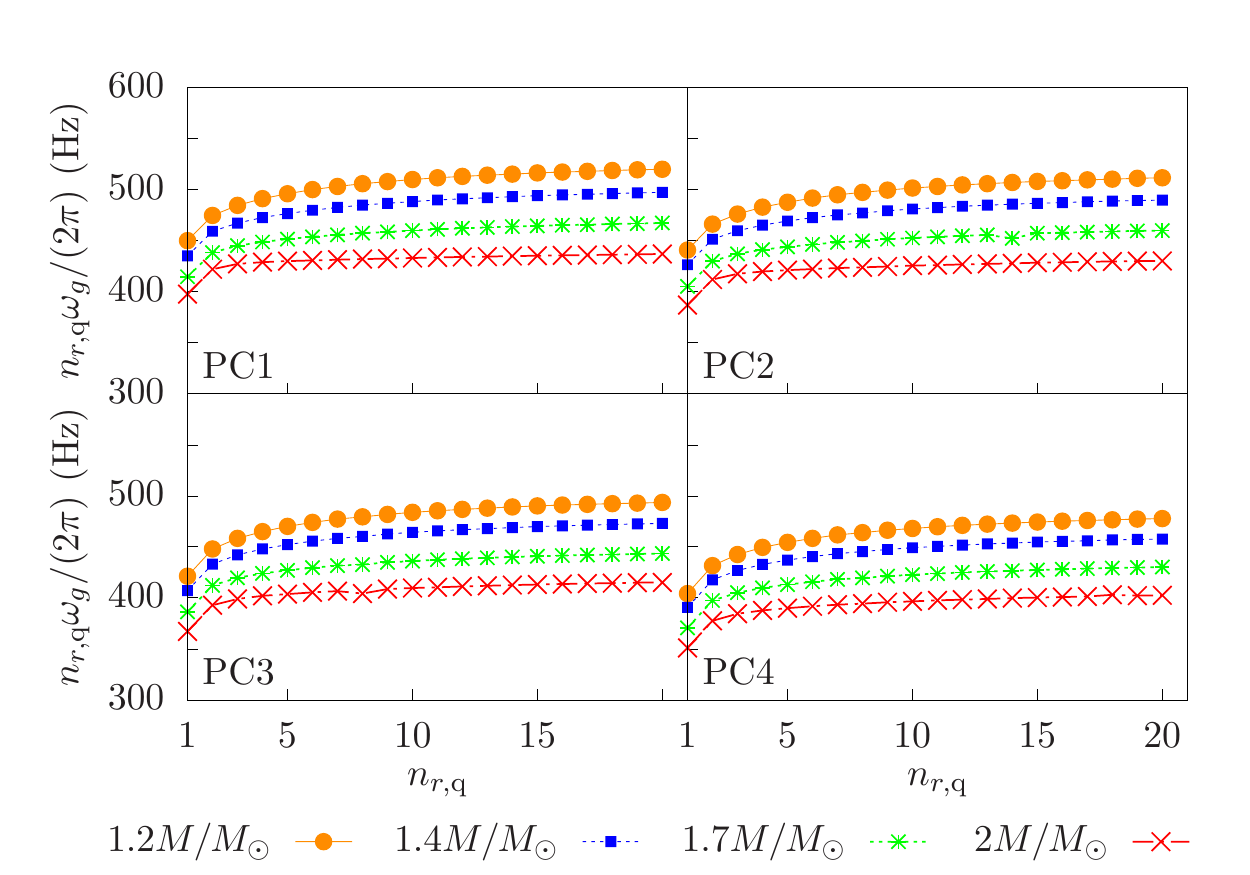}
\caption{$l=2$ $g$-mode (cyclical) frequencies for different values of the stellar mass and grouped by the EOS parametrization, denoted in the bottom left corner of each subplot. The entrainment in the core was set to zero when computing these frequencies.}
\label{fig:GModesMass}
\end{figure*}

\begin{figure*}
\centering
\includegraphics[width=1.0\textwidth]{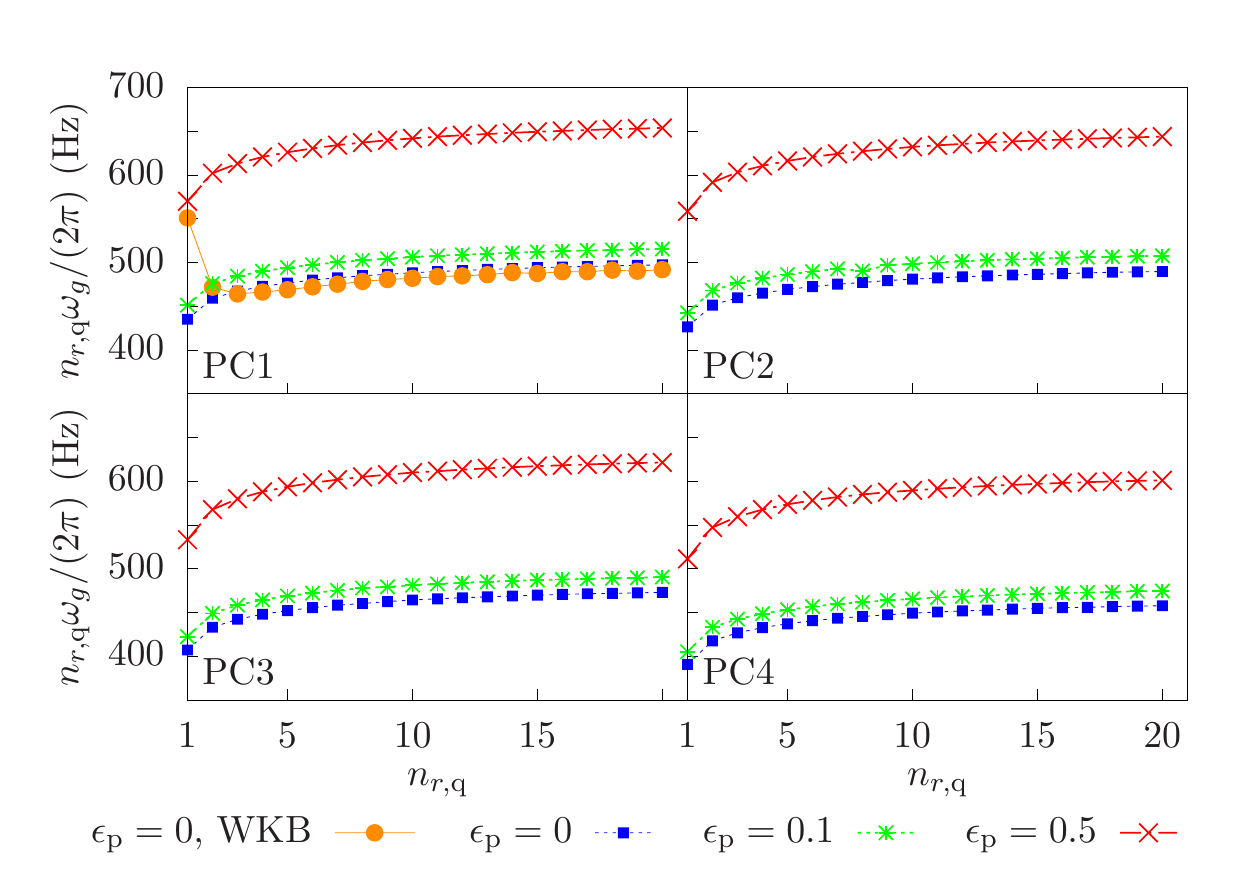}
\caption{$l=2$ $g$-mode (cyclical) frequencies for different values of the entrainment parameter $\epsilon_{\rm p}$, grouped by the EOS parametrization, denoted in the bottom left corner of each subplot. The WKB frequencies for a zero entrainment, $1.4M_{\odot}$ star with EOS parametrization PC1 is included in the upper left subplot. All stellar models used in this plot are of mass $1.4M_{\odot}$.}
\label{fig:GModesEntrainment}
\end{figure*}

Figures~\ref{fig:GModesMass} and~\ref{fig:GModesEntrainment} show the $l=2$ $g$-mode frequency spectrum as a function of the stellar mass and the entrainment parameter $\epsilon_{\rm p}$, respectively, for the four different EOS parametrizations described in Table~\ref{tab:EOSParameters}. The WKB frequencies for the $1.4M_{\odot}$ star with no entrainment are shown in Figure~\ref{fig:GModesEntrainment} to illustrate that they are extremely close to the exact frequencies for $n_{r,{\rm q}}\gtrsim 2$, even though we did not permit propagation into the crust in the WKB approximation. This indicates that the crust is largely unimportant to the mode frequency for the $g$-modes.

We find that our frequencies, which include general relativity, are redshifted compared to those of YW17. For example, they use the approximate $g$-mode frequency $\omega_g/2\pi\approx 590/n_{q,r}$ Hz for their $1.40M_{\odot}$, $m_{\rm p}^*/m_{\rm N}=0.8$ and $K=230.9$ MeV~\citep{RikovskaStone2003} star, while we obtain an approximate frequency spectrum
\begin{equation}
\frac{\omega_g}{2\pi}\approx\frac{608-0.83(K-240\text{ MeV})-90\frac{M}{M_{\odot}}+297\epsilon_{\rm p}}{n_{r,{\rm q}}}\text{ Hz}, \label{eq:OmegaGApprox}
\end{equation}
which is accurate to within $\lesssim$5\% for $n_{r,q}>2$. This formula gives $\omega_g/2\pi\approx 549/n_{r,{\rm q}}$ Hz for a $1.40M_{\odot}$, $\epsilon_{\rm p}=1-m_{\rm p}^*/m_{\rm N}=0.2$ and $K=230.9$ MeV star in our model. Our frequencies are also lower than those of KG14, who did include general relativity. In this case, the differences in the frequencies are due to the different equations of state used, which also contributed to the differences between the results in this paper and those in YW17. Eq.~\ref{eq:OmegaGApprox} also indicates that the $g$-mode frequency is relatively insensitive to the nuclear compressibility $K$, with the numerator changing by only $41$ Hz over the range of $K$ values used here.

As expected from the $(1+y)=(1-Y(1+\epsilon_{\rm p}))/(1-\epsilon_{\rm p}-Y)$ proportionality of the Brunt--V\"{a}is\"{a}l\"{a} frequency, the $g$-mode frequencies are increased as the entrainment parameter $\epsilon_{\rm p}$ is increased. We find that the frequencies increase by a factor or $\sim 1.4$ from the $\epsilon=0$ to the $\epsilon_{\rm p}=0.5$ values, in agreement with an expected scaling factor of $1/\sqrt{1-\epsilon_{\rm p}}=\sqrt{2}$ for low $Y$. However, we do not find an increase as large as found in YW17, with our frequencies with $\epsilon_{\rm p}=0.5$ being a factor of $\sim1.5$ lower than theirs with $m^*_{\rm p}/m_{\rm N}=0.4$. Possible reasons for the disagreement are the differences in the equation of state and in the structure of the star, which we compute by solving the TOV equation. The decrease in $\omega_g$ with $K$ is also expected from the inverse relationship between $N_{\rm q}$ and $K$ as seen in Figure~\ref{fig:BVFrequencyVarK}, though this decrease is small (hence the relative insensitivity to $K$) because the $g$-modes can propagate over a longer distance in higher $K$ stars due to their greater radii. The decrease in $\omega_g$ with $M$ , even though the maximum value of $N_{\rm q}$ in the star increases with $M$, is explained as follows. Figure~\ref{fig:BVFrequency} shows that $N_{\rm q}$ becomes more peaked as a function of mass, meaning that the region of the star where $k_r$ is real (between $r_t$ and $r_{\text{out}}$ in Eq.~(\ref{eq:krCondition})) is smaller for large $M$. For large $n_{r,q}$ Eq.~(\ref{eq:krCondition}) becomes
\begin{equation}
\omega_g \approx \frac{l}{n_{r,{\rm q}}\pi} \int_{r_t}^{r_{\text{out}}} \frac{N_{\rm q}}{r}dr,
\label{eq:ApproximatekrCondition}
\end{equation}
so $\omega_g$ is smaller for a particular $n_{r,{\rm q}}$ when the range of integration is smaller, or when $M$ is larger.

\begin{figure*}
\centering
\includegraphics[width=1.0\textwidth]{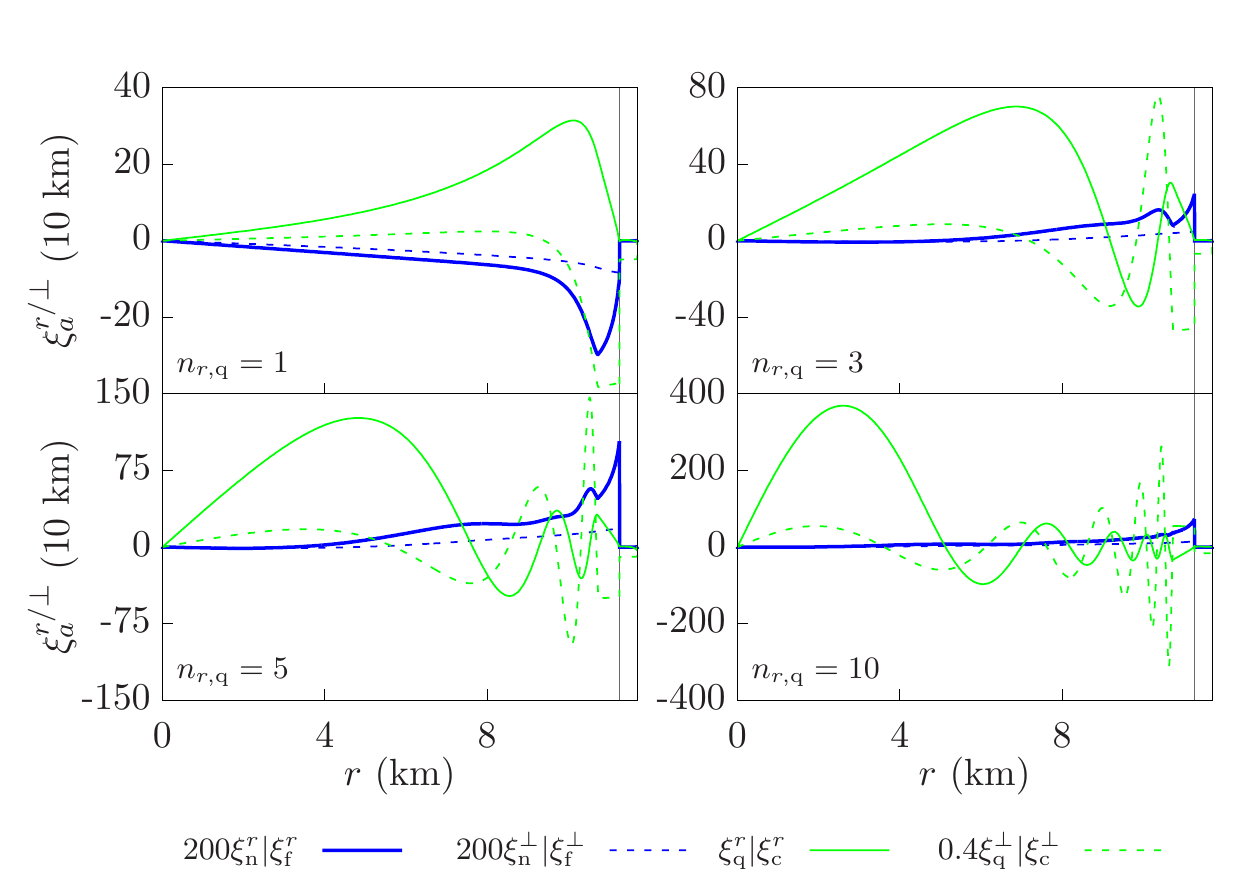}
\caption{Displacement fields $\xi^r$ and $\xi^{\perp}$ for four $l=2$ $g$-modes in a $1.40M_{\odot}$, zero entrainment star with EOS parametrization PC1: ($n_{r,{\rm q}}$, $\omega/2\pi$)=(1, 435.2 Hz), (3, 155.8 Hz), (5, 95.34 Hz) and (10, 48.84 Hz). The crust-core interface is indicated by the thin line at $11.27$ km. To the left of this line, the displacement fields are $(\xi^r_{\rm n},\xi^r_{\rm q},\xi^{\perp}_{\rm n},\xi^{\perp}_{\rm q})$, while to the right they are $(\xi^r_{\rm f},\xi^r_{\rm c},\xi^{\perp}_{\rm f},\xi^{\perp}_{\rm c})$. $(\xi^r_{\rm b},\xi^{\perp}_{\rm b})$, which do not vary much over the very thin region ($\sim 10$ m) of single fluid above neutron drip, are not shown.}
\label{fig:GModesDisplacementFields}
\end{figure*}

Figure~\ref{fig:GModesDisplacementFields} shows the displacement fields $\xi^r(r)$ and $\xi^{\perp}(r)$ for a few representative $l=2$ $g$-modes in the $1.40M_{\odot}$ star. Since the leptonic Brunt--V\"{a}is\"{a}l\"{a} frequency only acts on the charged fluid, the amplitude of the charged component is two orders of magnitude larger than the neutron component, and the neutron component is pulled along by the charge component through the thermodynamic coupling term $\mu_{\rm nq}$ (and also by the entrainment if $\epsilon_{\rm p}\neq 0$). In the crust, the $g$-mode frequencies are too low to excite oscillatory motion, and thus both the nuclear and neutron fluid displacements damp.

In the core, the charged component displacement fields are in reasonable agreement with YW17, but the neutron components have important differences. Our crust-core transition conditions change the oscillatory structure of the neutron component displacement fields, shifting them away from $\xi^r=0$ in the outer part of the core. This justifies our specification of the $g$-modes using $n_{r,{\rm q}}$, the radial node number of the charged fluid in the core. This is in contrast to results of YW17, which assumed a single normal fluid in the crust and imposes a crust-core transition condition (Eq. (B40) in YW17) that is equivalent to making both superfluid displacement fields equal. As the entrainment is increased in strength and the neutron fluid is forced to move along with the charged fluid to an even greater extent, we find that the radial nodes of the neutron fluid reappear at the locations of the charged fluid nodes. We cannot compare our results to YW17 in the crust because, unlike them, we treat the crust as two-fluids. We also do not compare our results to KG14, who do not show any displacement fields and who also use a single-fluid crust.


\subsubsection{$p$-modes}

\begin{figure*}
\centering
\includegraphics[width=1.0\textwidth]{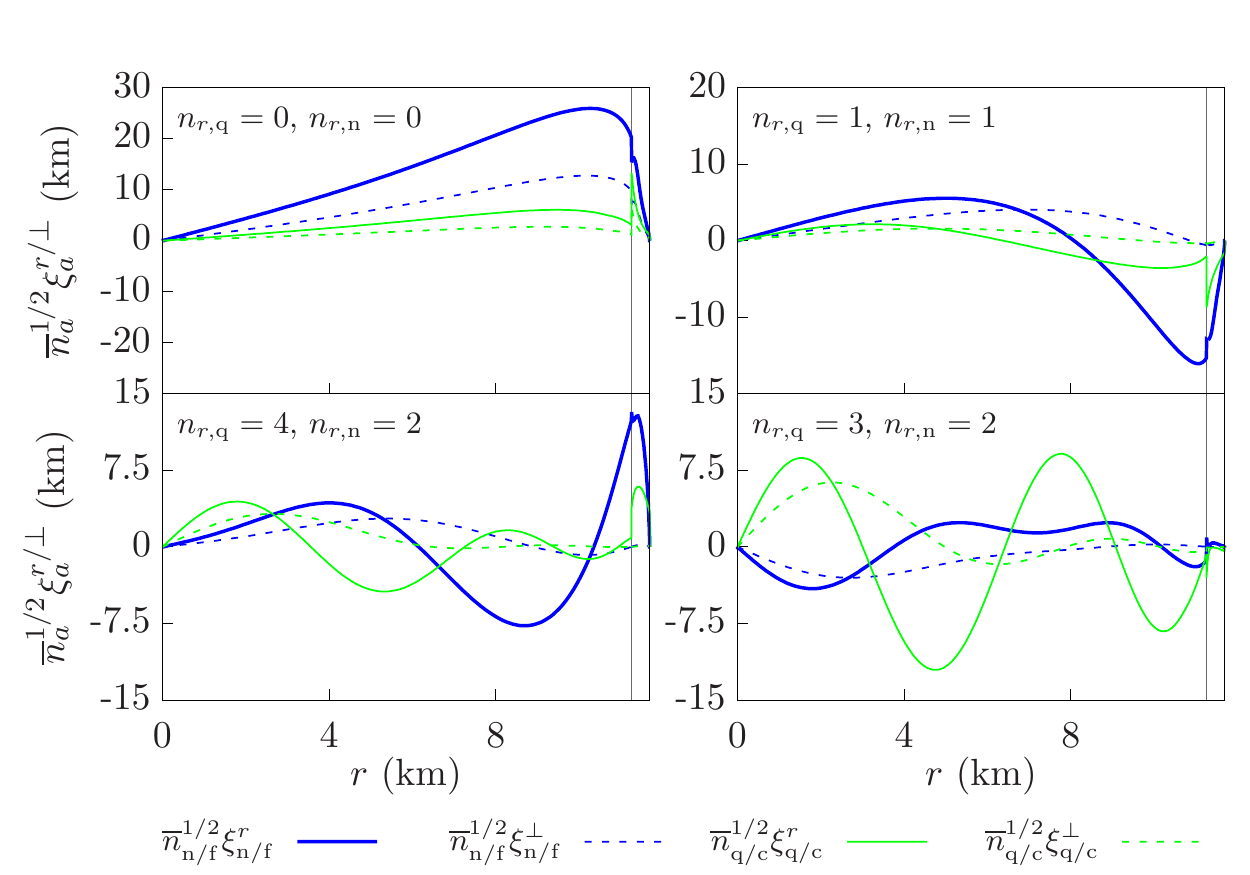}
\caption{Displacement fields $\xi^r$ and $\xi^{\perp}$ for the $l=2$ $f$-mode and three $p$-modes in a $1.40M_{\odot}$, zero entrainment star: ($x_i$, $n_{r,{\rm q}}$, $n_{r,{\rm n}}$, $\omega_g/2\pi$)=($f$, 0, 0, 2302 Hz), ($p_1$, 1, 1, 6576 Hz), ($p_2$, 4, 2, 9703 Hz) and ($s_3$, 3, 2, 10378 Hz), where $x_i$ refers to the standard classification of the mode and its order as a subscript. The crust-core interface is indicated by the thin line at $11.27$ km. To the left of this line, the displacement fields are $(\xi^r_{\rm n},\xi^r_{\rm q},\xi^{\perp}_{\rm n},\xi^{\perp}_{\rm q})$, while to the right they are $(\xi^r_{\rm f},\xi^r_{\rm c},\xi^{\perp}_{\rm f},\xi^{\perp}_{\rm c})$. $(\xi^r_{\rm b},\xi^{\perp}_{\rm b})$, which do not vary much over the very thin region ($\sim 10$ m) of single fluid above neutron drip, are not shown. The radial node numbers $n_{r,{\rm q}}$ and $n_{r,{\rm n}}$ for each fluid for each mode are indicated in the upper left of each plot. The displacement fields have been scaled by factors of $\overline{n}^{1/2}_a=(n_a/n_{\text{nuc}})^{1/2}$, which accounts for the abrupt jumps occurring at the crust-core transition.}
\label{fig:PModesDisplacementFields}
\end{figure*}

Figure~{\ref{fig:PModesDisplacementFields}} shows four distinct $l=2$ $p$-modes for a $\epsilon_{\rm p}$, $1.4M_{\odot}$ star, the first of which is actually an $n_{r,{\rm n}}=n_{r,{\rm q}}=0$ $f$-modes. These illustrate that 1) there are twice as many $p$-modes since there are two fluids, a result which is well-known \citep{Lindblom1994,Lee1995,Gualtieri2014}, including multiple modes with the same radial node number for one or both fluids, and 2) the fluids need not oscillate in phase, meaning the $n$ and $q$ fluids can have different numbers of radial nodes. In fact, we find that, for $\epsilon_{\rm p}=0$, most of the $p$-modes for a two-superfluid star behave as if the two fluids are (almost) uncoupled. This agrees with previous work \citep{Gusakov2011,Gualtieri2014}. This means that the core WKB result Eq.~(\ref{eq:ZerothOrderPhase}) does not apply for all $p$-modes since it assumes that the two fluids have identical phase, which is not necessarily true. In contrast to the $q$-led $g$-modes, the amplitudes of the $n$ and $q$-components of the $p$-modes are comparable. Additionally, the crust displacement fields can have multiple radial nodes, even with the crust constituting only a few percent of the star's radial extent, since the wave number for the $p$-mode is often significantly smaller in the crust than in the core. Following~\citet{Lindblom1994} we can classify $p$-modes by calculating the baryon current $Y\xi^r_{\rm q}+(1-Y)\xi^r_{\rm n}$: those with small baryon current compared with $\xi_{\rm n}^r-\xi_{\rm q}^r$  are classified as superfluid modes, denoted ``$s_i$'', while all others are classified as normal fluid modes, denoted ``$p_i$''. This is similar to the classification scheme of~\citet{Lindblom1994} and~\citet{Lee1995}, who use a scheme based on quantities related to our $\Pi_{\rm q}$ and $\Pi_{\rm n}$.
 
\begin{figure*}
\centering
\includegraphics[width=1.0\textwidth]{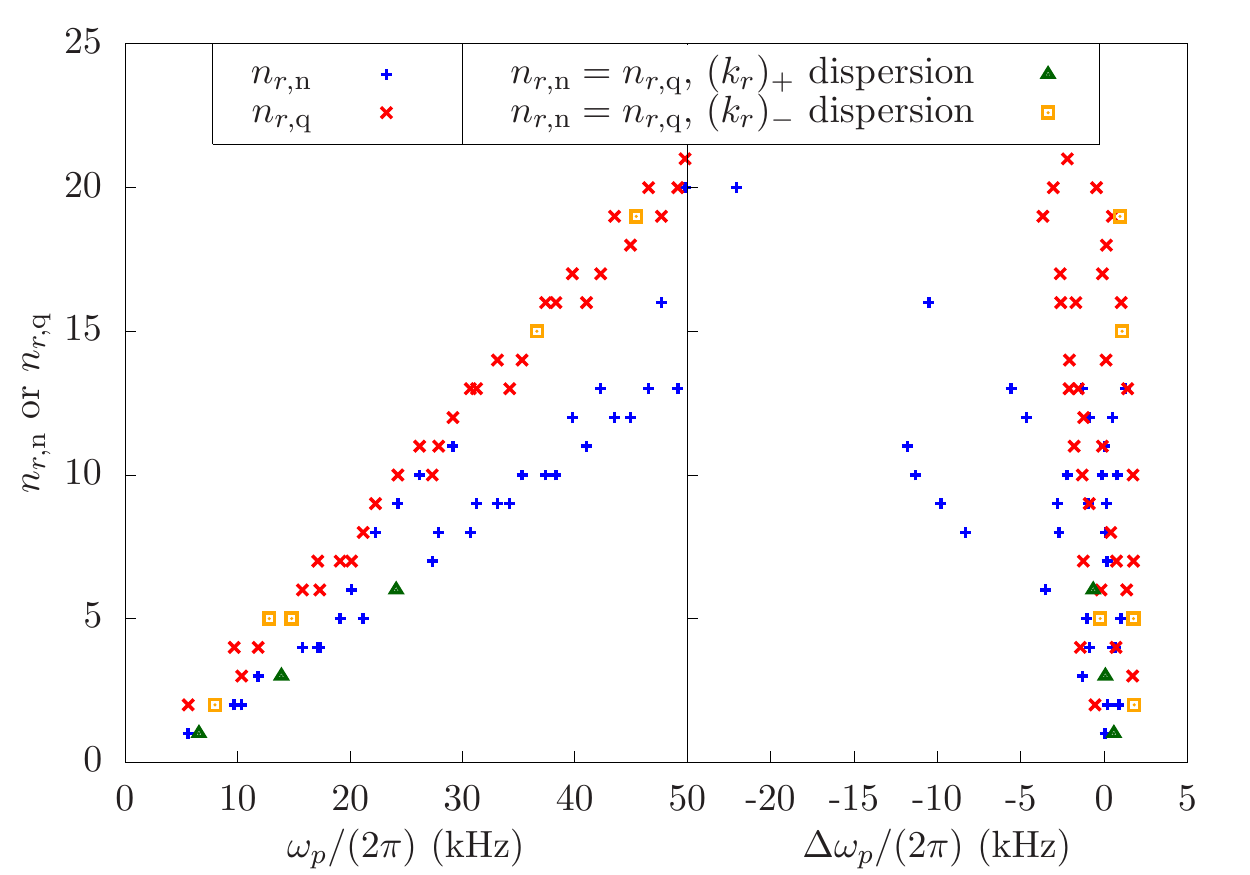}
\caption{Left: $l=2$ $p$-mode radial node numbers $n_{r,{\rm n}}$ and $n_{r,{\rm q}}$ plotted as a function of the frequency of the corresponding mode for a $1.40M_{\odot}$, zero entrainment star. The $f$ and $s_0$ ($\omega/2\pi=28687$ Hz) modes are not shown. Modes where $n_{r,{\rm n}}\neq n_{r,{\rm q}}$ have the radial node numbers of the $n$ and $q$ displacement fields  denoted separately, but are paired i.e. there are two ticks at the same frequency $\omega_{\rm p}$, one ($+$) denoting the value of $n_{r,{\rm n}}$ and the other (x) denoting $n_{r,{\rm q}}$. Modes where $n_{r,{\rm n}}=n_{r,{\rm q}}$ are denoted by distinct symbols depending on whether they more closely follow the $(k_r)_{+}$ ($n$-dominated, denoted by a triangle) or $(k_r)_{-}$ ($q$-dominated, denoted by a square) WKB dispersion relation. 
Right: Residuals $\Delta\omega_{\rm p}$ between the full numerically calculated $p$-mode frequencies and those that an uncoupled $n$ or $q$ mode of identical $n_{r,{\rm n}}$/$n_{r,{\rm q}}$ would have (for $n_{r,{\rm n}}\neq n_{r,{\rm q}}$) \textit{or} between the fully numerically calculated $p$-mode frequencies and the nearest coupled WKB frequency corresponding to the same radial node number $n_{r,{\rm n}}=n_{r,{\rm q}}$.}
\label{fig:PModesScatter}
\end{figure*}

Figure~\ref{fig:PModesScatter} plots the radial node numbers in the core for each $p$-mode as a function of the mode frequency with zero entrainment. We plot $n_{r,{\rm n}}$ and $n_{r,{\rm q}}$ separately for modes in which they are not identical and only one of them for modes in which they are the same. The $p$-modes for which $n_{r,{\rm n}}\neq n_{r,{\rm q}}$ the two components of the mode each roughly obey the uncoupled fluid dispersion relations $k^{\rm n}_r\approx M_{\rm nn}$ and $k^{\rm q}_r\approx M_{\rm qq}$, and those which have $n_{r,n}= n_{r,q}$ and roughly obey one of the two (coupled) WKB results given by Eq.~(\ref{eq:ZerothOrderPhase}). The modes of the latter type are labeled distinctly based on which solution $(k_r)_{\pm}$ they follow most closely. The separate, uncoupled dispersion relations obeyed by most $p$-modes suggest that they are formed from separate $n$ and $q$ oscillations which are paired together through the weak thermodynamic coupling (in the case of zero entrainment) due to having similar frequencies, with the pairing shifting the mode away from either of the exact frequencies that the uncoupled fluid modes would have. This means that the $n$ and $q$ components of each mode are not required to have the same node number, which is what we observe. The frequency residuals $\Delta\omega_{\rm p}$ compared to the uncoupled fluid or WKB result are shown in the right panel. For the nearly uncoupled modes, these were obtained by comparing the numerically calculated frequency to the expected frequency for the separate fluid components for a given $n_{r,{\rm n}}$ or $n_{r,{\rm q}}$ as calculated using $k^{\rm n}_r\approx M_{\rm nn}$ and $k^{\rm q}_r\approx M_{\rm qq}$ and Eq.~(\ref{eq:krCondition}). For the $n_{r,{\rm n}}=n_{r,{\rm q}}$ modes, the expected frequency was calculated for a given $n_r$ by using Eq.~(\ref{eq:krCondition}) and the WKB solution from Eq.~(\ref{eq:ZerothOrderPhase}) which gave the smallest frequency difference for each mode. $\Delta\omega_{\rm p}$ is small for most modes, indicating that they are well-described by either the nearly uncoupled or standard WKB dispersions. Many of the residuals for the high-frequency uncoupled-type neutron fluid modes are large, suggesting that for these modes the charged part of the mode could be ``pulling'' the neutron part towards being an $n_{r,{\rm n}}=n_{r,{\rm q}}$, charged fluid-like mode obeying the dispersion relation $(k_r)_{-}$.

\begin{figure}
\centering
\includegraphics[width=1.0\columnwidth]{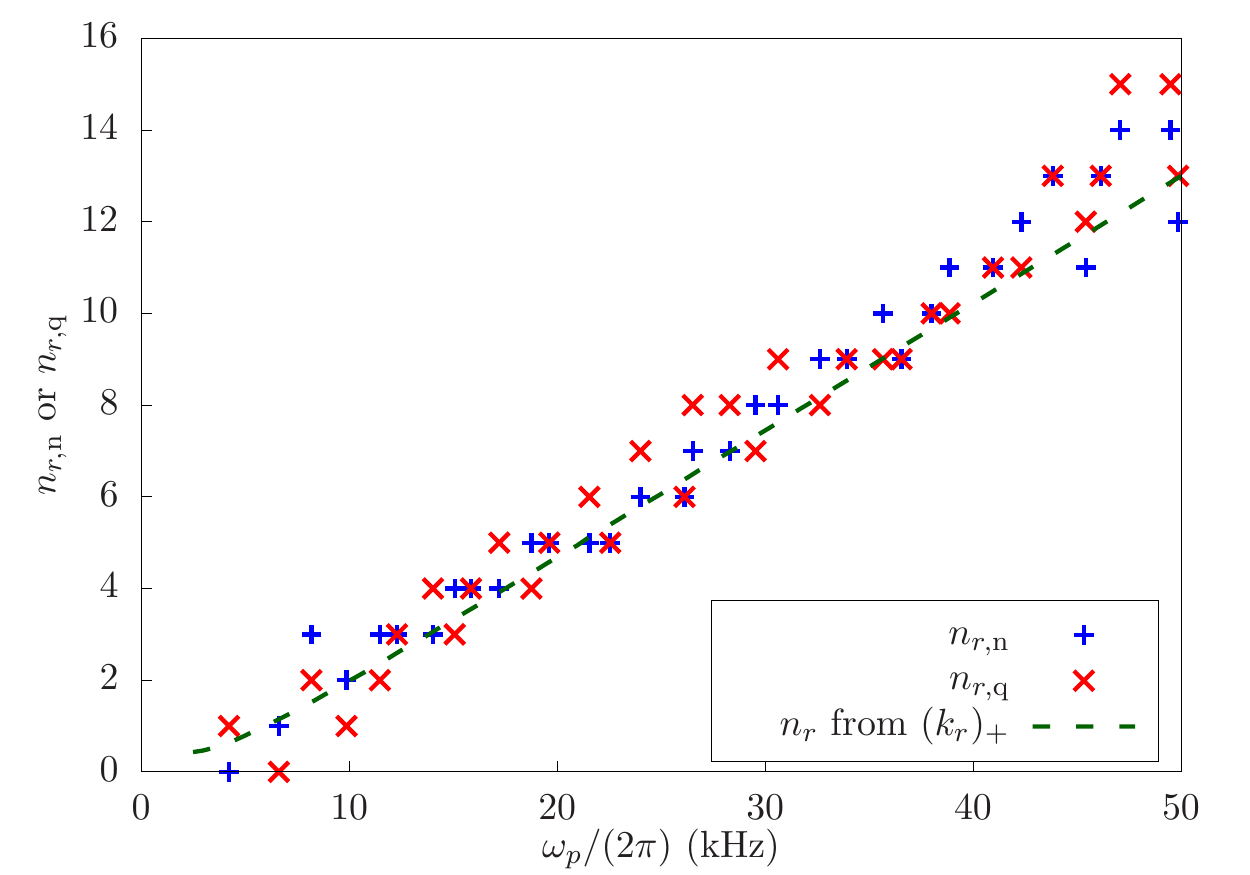}
\caption{$p$-modes for $1.40M_{\odot}$, $\epsilon_{\rm p}=0.5$ star with parametrization PC1 for the EOS, and $n_r$ (including fractional values) as a function of $\omega$ determined from Eq.~(\ref{eq:krCondition}) using $(k_r)_{+}\approx M_{\rm nn}$.}
\label{fig:PModesEntrainment}
\end{figure}

We also calculated the $p$-modes for a $1.40M_{\odot}$ star with strong entrainment $\epsilon_{\rm p}=0.5$. As expected, this drastic increase in the entrainment reduces the difference in the radial node number between the two fluids to at most $\pm 1$. It additionally tries to force the modes to obey the neutron-dominated WKB dispersion $(k_r)_{+}\approx M_{\rm nn}$, which is shown in  Figure~(\ref{fig:PModesEntrainment}). That it is this solution that is selected as opposed to the charged fluid-dominated one suggests that the $p$-modes can be thought of as neutron-dominated in the same way that the $g$-modes can be considered charge-fluid dominated, with this shift arising because the entrainment coefficient in the neutron equations $\epsilon_{\rm n}=n_{\rm q}/n_{\rm n}\epsilon_{\rm p}$ is about an order of magnitude smaller than $\epsilon_{\rm p}$, which appears in the charged fluid equations.

A final point of interest concerning the $p$-modes is the possible existence of pairs of distinct $p$-modes which are closely-spaced in frequency, as opposed to the nearly uniformly-spaced in frequency $p$-modes expected in the single fluid or WKB two-fluid cases. An example of such a mode pair we found for the $1.40M_{\odot}$, $K=230$ MeV, $\epsilon_{\rm p}=0$ star is the pair ($n_{r,{\rm q}}=6$, $n_{r,{\rm n}}=6$, $\omega_{\rm p}/2\pi=24116$ Hz) and ($n_{r,{\rm q}}=10$, $n_{r,{\rm n}}=9$, $\omega_{\rm p}/2\pi=24264$ Hz), which have a frequency spacing of the order of the $g$-mode frequencies. A similar phenomenon is observed in the finite temperature calculation of \citet{Gualtieri2014}, where the $p$-mode frequencies become very similar at certain ``resonance'' temperatures, though our results indicate that nearly-resonant $p$-modes can occur at any temperature. These mode pairs could provide a source of large nonlinear mode couplings for the two-superfluid version of the $p-g$ instability discussed in recent papers~\citep{Weinberg2013,Venumadhav2014,Weinberg2016}. These instabilities may be observable through phase shifts in the gravitational waveforms of binary neutron star mergers~\citep{Essick2016,Andersson2018}.

\section{Conclusions}

We have calculated the $g$- and $p$-modes of a superfluid star with leptonic buoyancy using a specific model for nuclear matter in the core and the crust. We have included general relativity and a two-fluid crust when computing the normal modes, finding that the crust-core interface conditions for the displacement fields change the neutron components of the $g$-mode displacement fields in the core by removing many of their radial nodes. In order to compute the modes, we have developed a simple but flexible equation of state for both crust and core which contains all of the thermodynamics required by our formalism. This allowed us to compute oscillation modes for a range of stellar and nuclear physics parameters, and our EOS can be easily adjusted to agree with new neutron star or nuclear physics measurements. In general our leptonic buoyancy $g$-mode frequencies are similar to those found previously, considering differences in the equations of state used to model the star and redshift factors, and are dominated by the charged fluid in which the buoyancy exists. We find that the $g$-mode frequencies increase with entrainment and decrease with stellar mass and nuclear compressibility, with only weak dependence on the latter. Our decomposition of the fluid into neutron and charged components clearly illustrates that the neutrons are pulled along by the charged fluid in the $g$-modes through thermodynamic coupling and entrainment, and otherwise would not participate in the $g$-mode.

In contrast, for zero entrainment, we reproduce earlier results~\citep{Gusakov2011,Gualtieri2014} that most of the $p$-modes behave as nearly uncoupled fluids, with the weak coupling between the two superfluids leading to pairing between uncoupled $n$- and $q$-fluid modes with similar frequencies. This results in $p$-modes whose charged and neutron components can have widely-differing radial node numbers, and in $p$-modes with frequency differences on the order of the $g$-mode frequencies. These could thus contribute to the recently proposed tidal-$p$-$g$ or related instabilities which depends on nonlinear couplings between $p$ and $g$-modes. For large entrainment, we find ``neutron-dominated'' $p$-modes, in which the phases of the two superfluids in the core are nearly the same so that they almost behave as a single neutron fluid.

As mentioned briefly by YW17 and incorporated in a recent paper~\citep{Yu2017a}, we should include hyperons in the neutron star core, as the chemical potential above $\sim 3n_{\text{nuc}}$ reaches the bare rest mass of the $\Lambda$ hyperon. This will lead to a softening of the equation of state and may provide additional hyperon superfluids which couple thermodynamically to the neutron and charged fluids, or if the hyperons are not superfluid, a hyperonic Brunt--V\"{a}is\"{a}l\"{a} frequency which can modify the $g$-modes in the inner core~\citep{Dommes2016}. If the star is able to contain $\Xi^{-}$ hyperons it could have a hyperonic Brunt--V\"{a}is\"{a}l\"{a} frequency even if the hyperons are superfluid, since the $\Xi^{-}$ would be expected to comove with the protons to which they are electrostatically coupled. Such hyperonic buoyancy would shift the $g$-mode frequencies obtained from leptonic buoyancy alone, which could be used  as an indicator of the presence of hyperons in neutron stars if the resulting gravitational waveform phase shifts from the resonant excitation of these $g$-modes in binary neutron star inspirals could be measured. However, if the EOS is softened too much by the hyperons, it could become difficult for it to allow stars of mass $>2M_{\odot}$, as reaching this mass already required large nuclear compressibilities or high central densities.

\section*{Acknowledgements}

This work was supported in part by NASA ATP grant NNX13AH42G. PBR was also supported in part by the Boochever Fellowship at Cornell for fall 2017. We also thank the referee for many helpful comments that improved our paper.

\bibliography{library,textbooks}

\begin{thebibliography}{}
\makeatletter
\relax
\def\mn@urlcharsother{\let\do\@makeother \do\$\do\&\do\#\do\^\do\_\do\%\do\~}
\def\mn@doi{\begingroup\mn@urlcharsother \@ifnextchar [ {\mn@doi@}
  {\mn@doi@[]}}
\def\mn@doi@[#1]#2{\def\@tempa{#1}\ifx\@tempa\@empty \href
  {http://dx.doi.org/#2} {doi:#2}\else \href {http://dx.doi.org/#2} {#1}\fi
  \endgroup}
\def\mn@eprint#1#2{\mn@eprint@#1:#2::\@nil}
\def\mn@eprint@arXiv#1{\href {http://arxiv.org/abs/#1} {{\tt arXiv:#1}}}
\def\mn@eprint@dblp#1{\href {http://dblp.uni-trier.de/rec/bibtex/#1.xml}
  {dblp:#1}}
\def\mn@eprint@#1:#2:#3:#4\@nil{\def\@tempa {#1}\def\@tempb {#2}\def\@tempc
  {#3}\ifx \@tempc \@empty \let \@tempc \@tempb \let \@tempb \@tempa \fi \ifx
  \@tempb \@empty \def\@tempb {arXiv}\fi \@ifundefined
  {mn@eprint@\@tempb}{\@tempb:\@tempc}{\expandafter \expandafter \csname
  mn@eprint@\@tempb\endcsname \expandafter{\@tempc}}}

\bibitem[\protect\citeauthoryear{Abbott et~al.,}{Abbott
  et~al.}{2017}]{Abbott2017}
Abbott B.,  et~al., 2017, \mn@doi [Phys. Rev. Lett.]
  {10.1103/PhysRevLett.119.161101}, 119, 161101

\bibitem[\protect\citeauthoryear{Agathos, Meidam, {Del Pozzo}, Li, Tompitak,
  Veitch, Vitale  \& {Van Den Broeck}}{Agathos et~al.}{2015}]{Agathos2015}
Agathos M.,  Meidam J.,  {Del Pozzo} W.,  Li T.~G.,  Tompitak M.,  Veitch J.,
  Vitale S.,   {Van Den Broeck} C.,  2015, \mn@doi [Phys. Rev. D]
  {10.1103/PhysRevD.92.023012}, 92, 023012

\bibitem[\protect\citeauthoryear{Andersson \& Comer}{Andersson \&
  Comer}{2007}]{Andersson2007}
Andersson N.,  Comer G.~L.,  2007, \mn@doi [Living Rev. Relativ.]
  {10.12942/lrr-2007-1}, 10, 1

\bibitem[\protect\citeauthoryear{Andersson \& Ho}{Andersson \&
  Ho}{2018}]{Andersson2018}
Andersson N.,  Ho W.~C.,  2018, \mn@doi [Phys. Rev. D]
  {10.1103/PhysRevD.97.023016}, 97, 23016

\bibitem[\protect\citeauthoryear{Andersson, Haskell  \& Samuelsson}{Andersson
  et~al.}{2011}]{Andersson2011}
Andersson N.,  Haskell B.,   Samuelsson L.,  2011, \mn@doi [Mon. Not. R.
  Astron. Soc.] {10.1111/j.1365-2966.2011.19015.x}, 416, 118

\bibitem[\protect\citeauthoryear{Antoniadis et~al.,}{Antoniadis
  et~al.}{2013}]{Antoniadis2013}
Antoniadis J.,  et~al., 2013, Science (80-. )., 340, 1233232

\bibitem[\protect\citeauthoryear{Baldo \& Schulze}{Baldo \&
  Schulze}{2007}]{Baldo2007}
Baldo M.,  Schulze H.~J.,  2007, \mn@doi [Phys. Rev. C]
  {10.1103/PhysRevC.75.025802}, 75, 025802

\bibitem[\protect\citeauthoryear{Baym, Pethick  \& Sutherland}{Baym
  et~al.}{1971a}]{Baym1971a}
Baym G.,  Pethick C.,   Sutherland P.,  1971a, \mn@doi [Astrophys. J.]
  {10.1086/151216}, 170, 299

\bibitem[\protect\citeauthoryear{Baym, Bethe  \& Pethick}{Baym
  et~al.}{1971b}]{Baym1971}
Baym G.,  Bethe H.~A.,   Pethick C.~J.,  1971b, \mn@doi [Nucl. Physics, Sect.
  A] {10.1016/0375-9474(71)90281-8}, 175, 225

\bibitem[\protect\citeauthoryear{Bertoni, Reddy  \& Rrapaj}{Bertoni
  et~al.}{2015}]{Bertoni2015}
Bertoni B.,  Reddy S.,   Rrapaj E.,  2015, \mn@doi [Phys. Rev. C]
  {10.1103/PhysRevC.91.025806}, 91, 025806

\bibitem[\protect\citeauthoryear{Bildsten \& Cutler}{Bildsten \&
  Cutler}{1992}]{Bildsten1992}
Bildsten L.,  Cutler C.,  1992, \mn@doi [Astrophys. J.] {10.1086/171983}, 400,
  175

\bibitem[\protect\citeauthoryear{Bildsten \& Cutler}{Bildsten \&
  Cutler}{1995}]{Bildsten1995}
Bildsten L.,  Cutler C.,  1995, \mn@doi [Astrophys. J.]
  {http://adsabs.harvard.edu/doi/10.1086/176099}, 449, 800

\bibitem[\protect\citeauthoryear{Carter \& Langlois}{Carter \&
  Langlois}{1998}]{Carter1998}
Carter B.,  Langlois D.,  1998, \mn@doi [Nucl. Phys. B]
  {10.1016/S0550-3213(98)00430-1}, 531, 478

\bibitem[\protect\citeauthoryear{Chamel}{Chamel}{2017}]{Chamel2017a}
Chamel N.,  2017, \mn@doi [J. Low Temp. Phys.] {10.1007/s10909-017-1815-x},
  189, 328

\bibitem[\protect\citeauthoryear{Cutler et~al.,}{Cutler
  et~al.}{1993}]{Cutler1993}
Cutler C.,  et~al., 1993, \mn@doi [Phys. Rev. Lett.]
  {https://doi.org/10.1103/PhysRevLett.70.2984}, 70, 2984

\bibitem[\protect\citeauthoryear{Dommes \& Gusakov}{Dommes \&
  Gusakov}{2016}]{Dommes2016}
Dommes V.~A.,  Gusakov M.~E.,  2016, \mn@doi [Mon. Not. R. Astron. Soc.]
  {10.1093/mnras/stv2408}, 455, 2852

\bibitem[\protect\citeauthoryear{Douchin \& Haensel}{Douchin \&
  Haensel}{2000}]{Douchin2000}
Douchin F.,  Haensel P.,  2000, \mn@doi [Phys. Lett. B]
  {https://doi.org/10.1016/S0370-2693(00)00672-9}, 485, 107

\bibitem[\protect\citeauthoryear{Epstein}{Epstein}{1988}]{Epstein1988a}
Epstein R.~I.,  1988, \mn@doi [Astrophys. J.] {10.1086/166797}, 333, 880

\bibitem[\protect\citeauthoryear{Essick, Vitale  \& Weinberg}{Essick
  et~al.}{2016}]{Essick2016}
Essick R.,  Vitale S.,   Weinberg N.~N.,  2016, \mn@doi [Phys. Rev. D]
  {https://journals.aps.org/prd/abstract/10.1103/PhysRevD.94.103012}, 94,
  103012

\bibitem[\protect\citeauthoryear{Flanagan \& Racine}{Flanagan \&
  Racine}{2007}]{Flanagan2007}
Flanagan {\'{E}}.~{\'{E}}.,  Racine {\'{E}}.,  2007, \mn@doi [Phys. Rev. D]
  {10.1103/PhysRevD.75.044001}, 75, 044001

\bibitem[\protect\citeauthoryear{Gezerlis \& Carlson}{Gezerlis \&
  Carlson}{2010}]{Gezerlis2010}
Gezerlis A.,  Carlson J.,  2010, \mn@doi [Phys. Rev. C]
  {10.1103/PhysRevC.81.025803}, 81, 025803

\bibitem[\protect\citeauthoryear{Gezerlis, Pethick  \& Schwenk}{Gezerlis
  et~al.}{2014}]{Gezerlis2014}
Gezerlis A.,  Pethick C.~J.,   Schwenk A.,  2014, in Bennemann K.~H.,
  Ketterson J.~B.,  eds, , Novel Superfluids, Volume 2.
Oxford University Press, Oxford, Chapt.~22, pp 580--616

\bibitem[\protect\citeauthoryear{Gualtieri, Kantor, Gusakov  \&
  Chugunov}{Gualtieri et~al.}{2014}]{Gualtieri2014}
Gualtieri L.,  Kantor E.~M.,  Gusakov M.~E.,   Chugunov A.~I.,  2014, \mn@doi
  [Phys. Rev. D] {10.1103/PhysRevD.90.024010}, 90, 024010

\bibitem[\protect\citeauthoryear{Gusakov \& Kantor}{Gusakov \&
  Kantor}{2011}]{Gusakov2011}
Gusakov M.~E.,  Kantor E.~M.,  2011, \mn@doi [Phys. Rev. D]
  {10.1103/PhysRevD.83.081304}, 83, 081304

\bibitem[\protect\citeauthoryear{Gusakov \& Kantor}{Gusakov \&
  Kantor}{2013}]{Gusakov2013a}
Gusakov M.~E.,  Kantor E.~M.,  2013, \mn@doi [Phys. Rev. D]
  {10.1103/PhysRevD.88.101302}, 88, 101302

\bibitem[\protect\citeauthoryear{Haensel}{Haensel}{2001}]{Haensel2001b}
Haensel P.,  2001, in Blaschke D.,  Glendenning N.~K.,   Sedrakian A.,  eds, ,
  Physics of Neutron Star Interiors.
Springer, Berlin, Chapt.~5, pp 127--174

\bibitem[\protect\citeauthoryear{Hashimoto, Seki  \& Yamada}{Hashimoto
  et~al.}{1984}]{Hashimoto1984}
Hashimoto M.,  Seki H.,   Yamada M.,  1984, \mn@doi [Prog. Theor. Phys.]
  {10.1143/PTP.71.320}, 71, 320

\bibitem[\protect\citeauthoryear{Hebeler, Lattimer, Pethick  \&
  Schwenk}{Hebeler et~al.}{2013}]{Hebeler2013}
Hebeler K.,  Lattimer J.~M.,  Pethick C.~J.,   Schwenk A.,  2013, \mn@doi
  [Astrophys. J.] {10.1088/0004-637X/773/1/11}, 773

\bibitem[\protect\citeauthoryear{Ho \& Lai}{Ho \& Lai}{1999}]{Ho1999}
Ho W. C.~G.,  Lai D.,  1999, \mn@doi [Mon. Not. R. Astron. Soc.]
  {https://doi.org/10.1046/j.1365-8711.1999.02703.x}, 308, 153

\bibitem[\protect\citeauthoryear{Kantor \& Gusakov}{Kantor \&
  Gusakov}{2014}]{Kantor2014}
Kantor E.~M.,  Gusakov M.~E.,  2014, \mn@doi [Mon. Not. R. Astron. Soc. Lett.]
  {10.1093/mnrasl/slu061}, 442, 90

\bibitem[\protect\citeauthoryear{Kobyakov \& Pethick}{Kobyakov \&
  Pethick}{2013}]{Kobyakov2013}
Kobyakov D.,  Pethick C.~J.,  2013, \mn@doi [Phys. Rev. C]
  {10.1103/PhysRevC.87.055803}, 87, 055803

\bibitem[\protect\citeauthoryear{Kobyakov \& Pethick}{Kobyakov \&
  Pethick}{2016}]{Kobyakov2016}
Kobyakov D.,  Pethick C.~J.,  2016, \mn@doi [Phys. Rev. C]
  {10.1103/PhysRevC.94.055806}, 94, 055806

\bibitem[\protect\citeauthoryear{Lackey \& Wade}{Lackey \&
  Wade}{2015}]{Lackey2015}
Lackey B.~D.,  Wade L.,  2015, \mn@doi [Phys. Rev. D]
  {10.1103/PhysRevD.91.043002}, 91, 043002

\bibitem[\protect\citeauthoryear{Lai}{Lai}{1994}]{Lai1994}
Lai D.,  1994, \mn@doi [Mon. Not. R. Astron. Soc.] {10.1093/mnras/270.3.611},
  270, 611

\bibitem[\protect\citeauthoryear{Lattimer \& Prakash}{Lattimer \&
  Prakash}{2016}]{Lattimer2016}
Lattimer J.~M.,  Prakash M.,  2016, \mn@doi [Phys. Rep.]
  {10.1016/j.physrep.2015.12.005}, 621, 127

\bibitem[\protect\citeauthoryear{Lattimer, Pethick, Ravenhall  \&
  Lamb}{Lattimer et~al.}{1985}]{Lattimer1985}
Lattimer J.~M.,  Pethick C.~J.,  Ravenhall D.~G.,   Lamb D.,  1985, \mn@doi
  [Nucl. Phys. A] {https://doi.org/10.1016/0375-9474(85)90006-5}, 432, 646

\bibitem[\protect\citeauthoryear{Lee}{Lee}{1995}]{Lee1995}
Lee U.,  1995, \mn@doi [Astron. Astrophys.] {1995A&A...303..515L}, 303, 515

\bibitem[\protect\citeauthoryear{Lindblom \& Mendell}{Lindblom \&
  Mendell}{1994}]{Lindblom1994}
Lindblom L.,  Mendell G.,  1994, \mn@doi [Astrophys. J.] {10.1086/173682}, 421,
  689

\bibitem[\protect\citeauthoryear{Lombardo \& Schulze}{Lombardo \&
  Schulze}{2001}]{Lombardo2001}
Lombardo U.,  Schulze H.-J.,  2001, in Blaschke D.,  Glendenning N.~K.,
  Sedrakian A.,  eds, , Physics of Neutron Star Interiors.
Springer, Berlin, Chapt.~2, pp 30--53

\bibitem[\protect\citeauthoryear{McDermott, {Van Horn}  \& Scholl}{McDermott
  et~al.}{1983}]{McDermott1983}
McDermott P.,  {Van Horn} H.,   Scholl J.,  1983, \mn@doi [Astrophys. J.]
  {10.1086/161006}, 268, 837

\bibitem[\protect\citeauthoryear{Onsi, Dutta, Chatri, Goriely, Chamel  \&
  Pearson}{Onsi et~al.}{2008}]{Onsi2008}
Onsi M.,  Dutta A.~K.,  Chatri H.,  Goriely S.,  Chamel N.,   Pearson J.~M.,
  2008, \mn@doi [Phys. Rev. C] {10.1103/PhysRevC.77.065805}, 77, 062805

\bibitem[\protect\citeauthoryear{Page, Prakash, Lattimer  \& Steiner}{Page
  et~al.}{2011}]{Page2011}
Page D.,  Prakash M.,  Lattimer J.~M.,   Steiner A.~W.,  2011, \mn@doi [Phys.
  Rev. Lett.] {10.1103/PhysRevLett.106.081101}, 106, 081101

\bibitem[\protect\citeauthoryear{Passamonti \& Andersson}{Passamonti \&
  Andersson}{2012}]{Passamonti2012}
Passamonti A.,  Andersson N.,  2012, \mn@doi [Mon. Not. R. Astron. Soc.]
  {10.1111/j.1365-2966.2011.19725.x}, 419, 638

\bibitem[\protect\citeauthoryear{Passamonti, Andersson  \& Ho}{Passamonti
  et~al.}{2016}]{Passamonti2016}
Passamonti A.,  Andersson N.,   Ho W. C.~G.,  2016, \mn@doi [Mon. Not. R.
  Astron. Soc.] {10.1093/mnras/stv2149}, 455, 1489

\bibitem[\protect\citeauthoryear{Pearson, Chamel, Goriely  \& Ducoin}{Pearson
  et~al.}{2012}]{Pearson2012}
Pearson J.~M.,  Chamel N.,  Goriely S.,   Ducoin C.,  2012, \mn@doi [Phys. Rev.
  C] {10.1103/PhysRevC.85.065803}, 85, 065803

\bibitem[\protect\citeauthoryear{Potekhin, Fantina, Chamel, Pearson  \&
  Goriely}{Potekhin et~al.}{2013}]{Potekhin2013}
Potekhin A.~Y.,  Fantina A.~F.,  Chamel N.,  Pearson J.~M.,   Goriely S.,
  2013, \mn@doi [Astron. Astrophys.] {10.1051/0004-6361/201321697}, 560, A48

\bibitem[\protect\citeauthoryear{Prix \& Rieutord}{Prix \&
  Rieutord}{2002}]{Prix2002}
Prix R.,  Rieutord M.,  2002, \mn@doi [Astron. Astrophys.]
  {10.1051/0004-6361:20021049}, 393, 949

\bibitem[\protect\citeauthoryear{Ravenhall, Bennett  \& Pethick}{Ravenhall
  et~al.}{1972}]{Ravenhall1972}
Ravenhall D.~G.,  Bennett C.~D.,   Pethick C.~J.,  1972, \mn@doi [Phys. Rev.
  Lett.] {https://doi.org/10.1103/PhysRevLett.28.978}, 28, 978

\bibitem[\protect\citeauthoryear{Ravenhall, Pethick  \& Wilson}{Ravenhall
  et~al.}{1983a}]{Ravenhall1983}
Ravenhall D.~G.,  Pethick C.~J.,   Wilson J.~R.,  1983a, \mn@doi [Phys. Rev.
  Lett.] {https://doi.org/10.1103/PhysRevLett.50.2066}, 50, 2066

\bibitem[\protect\citeauthoryear{Ravenhall, Pethick  \& Lattimer}{Ravenhall
  et~al.}{1983b}]{Ravenhall1983a}
Ravenhall D.~G.,  Pethick C.~J.,   Lattimer J.~M.,  1983b, \mn@doi [Nucl. Phys.
  A] {https://doi.org/10.1016/0375-9474(83)90667-X}, 407, 571

\bibitem[\protect\citeauthoryear{Reisenegger \& Goldreich}{Reisenegger \&
  Goldreich}{1992}]{Reisenegger1992}
Reisenegger A.,  Goldreich P.,  1992, \mn@doi [Astrophys. J.] {10.1086/171645},
  395, 240

\bibitem[\protect\citeauthoryear{Reisenegger \& Goldreich}{Reisenegger \&
  Goldreich}{1994}]{Reisenegger1994}
Reisenegger A.,  Goldreich P.,  1994, \mn@doi [Astrophys. J.] {10.1086/174105},
  426, 688

\bibitem[\protect\citeauthoryear{{Rikovska Stone}, Miller, Koncewicz, Stevenson
   \& Strayer}{{Rikovska Stone} et~al.}{2003}]{RikovskaStone2003}
{Rikovska Stone} J.,  Miller J.~C.,  Koncewicz R.,  Stevenson P.~D.,   Strayer
  M.~R.,  2003, \mn@doi [Phys. Rev. C] {10.1103/PhysRevC.68.034324}, 68, 16

\bibitem[\protect\citeauthoryear{Venumadhav, Zimmerman  \& Hirata}{Venumadhav
  et~al.}{2014}]{Venumadhav2014}
Venumadhav T.,  Zimmerman A.,   Hirata C.~M.,  2014, \mn@doi [Astrophys. J.]
  {10.1088/0004-637X/781/1/23}, 781, 23

\bibitem[\protect\citeauthoryear{Watanabe, Sato, Yasuoka  \&
  Ebisuzaki}{Watanabe et~al.}{2003}]{Watanabe2003}
Watanabe G.,  Sato K.,  Yasuoka K.,   Ebisuzaki T.,  2003, \mn@doi [Phys. Rev.
  C] {10.1103/PhysRevC.68.035806}, 68, 21

\bibitem[\protect\citeauthoryear{Weinberg}{Weinberg}{2016}]{Weinberg2016}
Weinberg N.~N.,  2016, \mn@doi [Astrophys. J.] {10.3847/0004-637X/819/2/109},
  819, 109

\bibitem[\protect\citeauthoryear{Weinberg, Arras  \& Burkart}{Weinberg
  et~al.}{2013}]{Weinberg2013}
Weinberg N.~N.,  Arras P.,   Burkart J.,  2013, \mn@doi [Astrophys. J.]
  {10.1088/0004-637X/769/2/121}, 769, 121

\bibitem[\protect\citeauthoryear{Xu \& Lai}{Xu \& Lai}{2017}]{Xu2017}
Xu W.,  Lai D.,  2017, \mn@doi [Phys. Rev. D] {10.1103/PhysRevD.96.083005}, 96,
  083005

\bibitem[\protect\citeauthoryear{Yakovlev, Levenfish  \& Shibanov}{Yakovlev
  et~al.}{1999}]{Yakovlev1999}
Yakovlev D.~G.,  Levenfish K.~P.,   Shibanov Y.~A.,  1999, Rev. Top. Probl.,
  42, 737

\bibitem[\protect\citeauthoryear{Yu \& Weinberg}{Yu \&
  Weinberg}{2017a}]{Yu2017}
Yu H.,  Weinberg N.~N.,  2017a, \mn@doi [Mon. Not. R. Astron. Soc.]
  {10.1093/mnras/stw2552}, 464, 2622

\bibitem[\protect\citeauthoryear{Yu \& Weinberg}{Yu \&
  Weinberg}{2017b}]{Yu2017a}
Yu H.,  Weinberg N.~N.,  2017b, \mn@doi [Mon. Not. R. Astron. Soc.]
  {10.1093/mnras/stx1188}, 470, 350

\bibitem[\protect\citeauthoryear{Zhou, Schulze, Zhao, Pan  \& Draayer}{Zhou
  et~al.}{2004}]{Zhou2004}
Zhou X.~R.,  Schulze H.~J.,  Zhao E.~G.,  Pan F.,   Draayer J.~P.,  2004,
  \mn@doi [Phys. Rev. C] {10.1103/PhysRevC.70.048802}, 70, 048802

\makeatother
\end{thebibliography}

\appendix
\section{Calculation of chemical potential derivatives in the crust}
\label{app:1}

Computing $\mu_{\rm ff}$, $\mu_{\rm cc}$ and $\mu_{\rm fc}$ is more complicated than finding the $\mu_{ab}$ in the core. In the core, the energy density is a function of three variables, chosen to be $(n_{\rm n},n_{\rm q},f)$ or $(n_{\rm b},Y=n_{\rm q}/n_{\rm b},f)$. In the background, chemical equilibrium relates $Y$ and $f$ to $n_{\rm b}$; in perturbations, chemical equilibrium fails. In the crust, the energy density depends on five variables, three of which describe nuclei: $(A,n_{\rm i},Y)$. In equilibrium, there are four conditions that permit computing $A$, $n_{\rm i}$, $Y$ and $n_{\rm c}$ (or $Y_{\rm c}\equiv n_{\rm c}/n_{\rm b}$) as functions of $n_{\rm b}$; for perturbations, there are still three conditions, which may be solved in principle to find $A(n_{\rm f},n_{\rm c})$, $n_{\rm i}(n_{\rm f},n_{\rm c})$ and $Y(n_{\rm f},n_{\rm c})$, which we would then need to differentiate to compute $\mu_{ab}$ $a,b\in\{\rm f,c\}$.

Fortunately, perturbations away from equilibrium are small, so we only need to find the energy density $\rho$ as a function of $(n_{\rm b},Y_{\rm c})$ near equilibrium to compute $\mu_{ab}$. Define
\begin{equation}
\Delta Y_{\rm c} = Y_{\rm c}-Y_{\rm c}^{\text{eq}}(n_{\rm b}),
\end{equation}
where $Y_{\rm c}^{\text{eq}}(n_{\rm b})$ is the equilibrium value of $Y_{\rm c}$ and $\Delta Y_{\rm c} = 0$ in equilibrium. Then near but not in equilibrium, the energy density is
\begin{equation}
\rho(n_{\rm b},Y_{\rm c})=\rho_{\text{eq}}(n_{\rm b}) +\frac{1}{2}C(n_{\rm b})(\Delta Y_{\rm c})^2,
\label{eq:CrustEnergyDensityAppendixForm}
\end{equation}
where $\rho_{\text{eq}}(n_{\rm b})$ is evaluated in the background and
\begin{equation}
C(n_{\rm b})(\Delta Y_{\rm c})^2=\sum_{i,j}\left(\frac{\partial^2\rho(n_{\rm b},Y_{\rm c},A,Y,n_{\rm i})}{\partial X_i\partial X_j}\right)\Delta X_i\Delta X_j, \label{eq:EnergyDensityExpansion}
\end{equation}
where $X_i=A,Y,n_{\rm i},Y_{\rm c}$.

\begin{figure}
\centering
\includegraphics[width=1.0\columnwidth]{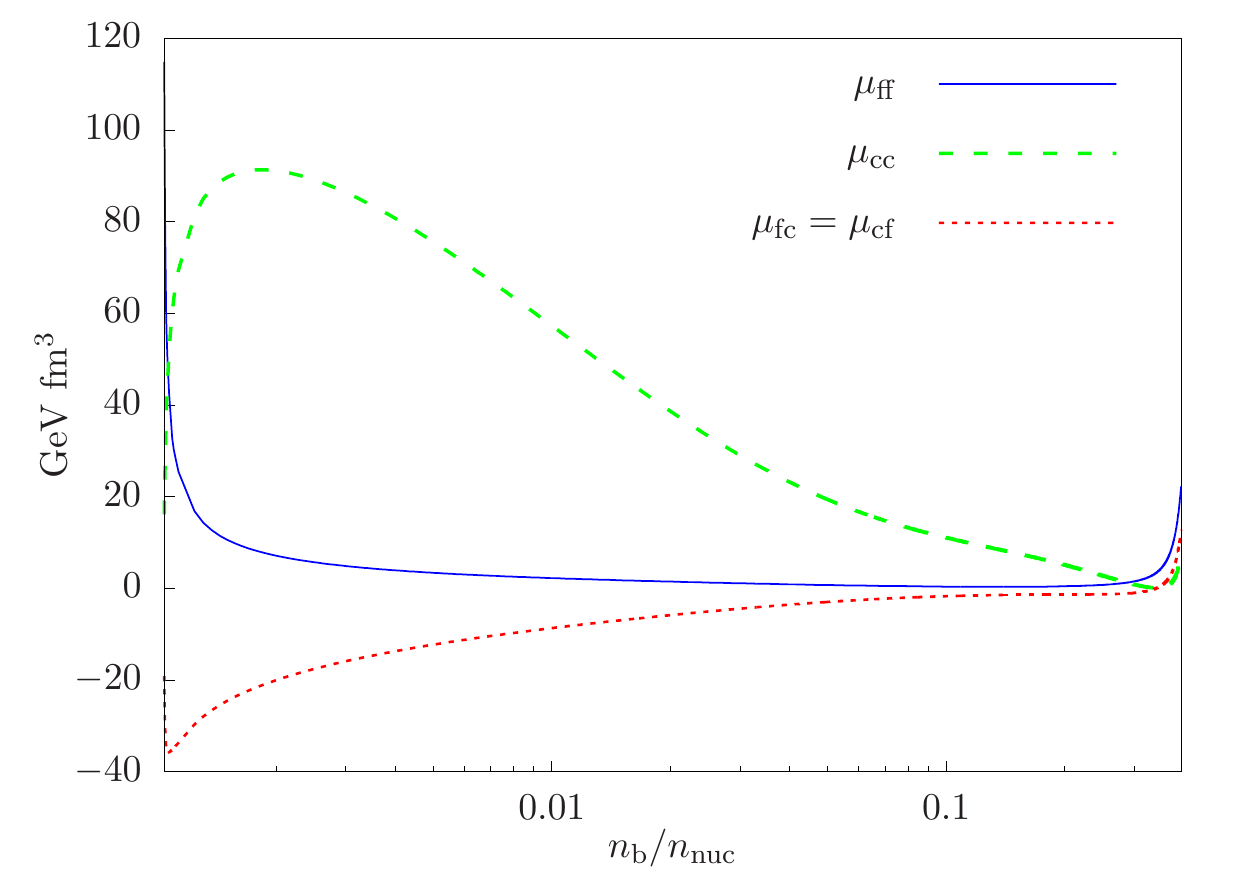}
\caption{Thermodynamic derivatives in the crust $\mu_{\rm ff}$, $\mu_{\rm cc}$ and $\mu_{\rm fc}$ as a function of the baryon density $0.00104<n_{\rm b}/n_{\text{nuc}}<0.394$ for the PC1 parametrization of our EOS.}
\label{fig:CrustMuxx}
\end{figure}

Perturbed fluid elements will quickly reach mechanical equilibrium and partial chemical equilibrium (i.e. not beta equilibrium) with their surroundings and will obey the ``nuclear virial theorem''. These conditions are expressed as
\begin{align}
&E_{\text{Coul}}=2E_{\text{surf}}, \label{eq:NuclearVirialTheorem} \\
&\mu_{\rm n,i}-\mu_{\rm n,o}=\frac{4\pi r_{\rm n}^2 Y}{A}\frac{{\rm d}\sigma_{\rm s}}{{\rm d}Y}, \label{eq:CrustPressureBalance} \\
&P_{{\rm i},\text{bulk}}-P_{{\rm o},\text{bulk}}=\frac{2\sigma_{\rm s}}{r_{\rm n}}-\frac{4\pi}{15}(Yn_{\rm i}r_{\rm n}e)^2(1-w), \label{eq:CrustExchangeCondition}
\end{align} 
where we have defined
\begin{align}
\mu_{\rm n,i}\equiv{}&\frac{\partial\rho_{\text{bulk},{\rm i}}}{\partial(n_{\rm i}(1-Y))}, \\
\mu_{\rm n,o}\equiv{}&\frac{\partial\rho_{\text{bulk},{\rm o}}}{\partial n_{\rm n,o}}, \\
P_{{\rm i},\text{bulk}}\equiv{}& n_{\rm i}^2\frac{\partial}{\partial n_{\rm i}}\left(\frac{\rho_{\text{bulk},{\rm i}}}{n_{\rm i}}\right), \\
P_{{\rm o},\text{bulk}}\equiv{}& n_{\rm n,o}^2\frac{\partial}{\partial n_{\rm n,o}}\left(\frac{\rho_{\text{bulk},{\rm o}}}{n_{\rm n,o}}\right).
\end{align}
The perturbed fluid elements will not be in beta equilibrium with their surroundings, since the weak interaction timescale is much longer than the timescale of the fluid oscillations, so $\partial\rho/\partial Y\neq 0$.
We then use Eqs.~(\ref{eq:NuclearVirialTheorem}--\ref{eq:CrustExchangeCondition}) to relate $\Delta Y_{\rm c}$ to $\Delta A$, $\Delta Y$ and $\Delta n_{\rm i}$ as in Eq.~(\ref{eq:EnergyDensityExpansion}) taking the differential of these three equations gives
\begin{align}
0={}&\rho_{AA}\Delta A+\rho_{An_{\rm i}}\Delta n_{\rm i}+\rho_{AY}\Delta Y + \rho_{AY_{\rm c}}\Delta Y_{\rm c}, \\
0={}&\rho_{n_{\rm i}n_{\rm i}}\Delta n_{\rm i}+\rho_{An_{\rm i}}\Delta A+\rho_{Yn_{\rm i}}\Delta Y + \rho_{n_{\rm i}Y_{\rm c}}\Delta Y_{\rm c}, \\
0={}&\rho_{Y_{\rm c}Y_{\rm c}}\Delta Y_{\rm c}+\rho_{n_{\rm i}Y_{\rm c}}\Delta n_{\rm i}+\rho_{AY_{\rm c}}\Delta A + \rho_{YY_{\rm c}}\Delta Y,
\end{align}
where $\rho_{X_iX_j}\equiv\partial^2\rho/(\partial X_i\partial X_j)$. Eq.~(\ref{eq:CrustEnergyDensityAppendixForm}) then gives the thermodynamic derivatives $\mu_{\rm bb}$, $\mu_{Y_{\rm c}Y_{\rm c}}$ and $\mu_{{\rm b}Y_{\rm c}}=\mu_{Y_{\rm c}{\rm b}}$ as
\begin{align}
\mu_{\rm bb}={}&\frac{d^2\rho_{\text{eq}}}{dn_{\rm b}^2}+C(n_{\rm b})\left(\frac{{\rm d}Y_{\rm c}^{\text{eq}}}{{\rm d}n_{\rm b}}\right)^2, \\
\mu_{Y_{\rm c}Y_{\rm c}}={}&C(n_{\rm b}), \\
\mu_{{\rm b}Y_{\rm c}}={}&-C(n_{\rm b})\frac{{\rm d}Y_{\rm c}^{\text{eq}}}{{\rm d}n_{\rm b}},
\end{align}
using which $\mu_{\rm cc}$, $\mu_{\rm ff}$ and $\mu_{\rm fc}=\mu_{\rm cf}$ are found using Eqs.~(\ref{eq:Munn}--\ref{eq:Munq}), replacing $Y\rightarrow Y_{\rm c}$, $q\rightarrow c$ and $n\rightarrow f$. The expressions for $\mu_{\rm cc}$, $\mu_{\rm ff}$ and $\mu_{\rm fc}$ are quite complicated and are not given explicitly here, but they are plotted in Figure~\ref{fig:CrustMuxx} for the PC1 parametrization of our EOS.

\bsp	
\label{lastpage}
\end{document}